\documentclass[10pt,journal,comsoc]{IEEEtran}
\usepackage{graphicx}
\usepackage{amsmath}
\usepackage{comment}

\ifCLASSOPTIONcompsoc
  \usepackage[nocompress]{cite}
\else
  \usepackage{cite}
\fi
\ifCLASSINFOpdf
\else
\fi
\begin{document}
%\title{Bare Demo of IEEEtran.cls for\\ IEEE Computer Society Journals}
\title{Network Centralities in Quantum Entanglement Distribution due to User Preferences}

\author{Dibakar~Das,~\IEEEmembership{Senior~Member,~IEEE,}
        Shiva~Kumar~Malapaka,
        Jyotsna~Bapat,~\IEEEmembership{Member,~IEEE,}
        and~Debabrata~Das,~\IEEEmembership{Senior~Member,~IEEE}% <-this % stops a space
\IEEEcompsocitemizethanks{\IEEEcompsocthanksitem
Dibakar Das is a researcher at IIIT Bangalore, Electronics City, Bangalore - 560100.\protect\\
E-mail: dibakard@ieee.org
\IEEEcompsocthanksitem         Shiva~Kumar~Malapaka,
        Jyotsna~Bapat, and Debabrata~Das are faculties at IIIT Bangalore.}% <-this % stops an unwanted space
\thanks{Manuscript received TBD; revised August TBD.}}


\markboth{Journal of \LaTeX\ Class Files,~Vol.~14, No.~8, August~2015}%
{Shell \MakeLowercase{\textit{et al.}}: Bare Demo of IEEEtran.cls for Computer Society Journals}



\IEEEtitleabstractindextext{%
\begin{abstract}
Quantum networks are of great interest of late which apply quantum mechanics to transfer information securely. One of the key properties which are exploited is entanglement to transfer information from one network node to another. Applications like quantum teleportation rely on the entanglement between the concerned nodes. Thus, efficient entanglement distribution among network nodes is of utmost importance. Several entanglement distribution methods have been proposed in the literature which primarily rely on attributes, such as, fidelities, link layer network topologies, proactive distribution, etc.  This paper studies the centralities of the network when the link layer topology of entanglements (referred to as entangled graph) is driven by usage patterns of peer-to-peer connections between remote nodes (referred to as connection graph) with different characteristics. Three different distributions (uniform, gaussian, and power law) are considered for the connection graph where the two nodes are selected from the same distribution. For the entangled graph, both reactive and proactive entanglements are employed to form a random graph. Results show that the edge centralities (measured as usage frequencies of individual edges during entanglement distribution) of the entangled graph follow power law distributions whereas the growth in entanglements with connections and node centralities (degrees of nodes) are monomolecularly distributed for most of the scenarios. These findings will help in quantum resource management, e.g., quantum technology with high reliability and lower decoherence time may be allocated to edges with high centralities.
\end{abstract}

\begin{IEEEkeywords}
centrality, edge, node, quantum, networks, power law, monomolecular
\end{IEEEkeywords}}

\maketitle

\IEEEdisplaynontitleabstractindextext

% \fi
\IEEEpeerreviewmaketitle

%\IEEEraisesectionheading{\section{Introduction}\label{sec:introduction}}
\section{Introduction}\label{section_introduction}

%\IEEEPARstart{T}{his} demo file is intended to serve as a ``starter file''
% for IEEE Computer Society journal papers produced under \LaTeX\ using
% IEEEtran.cls version 1.8b and later.
% I wish you the best of success.
\IEEEPARstart{Q}{uantum} networks are of great research interest in recent years \cite{cite_quantum_internet_protocol_survey}\cite{cite_quantum_internet_survey_comsoc}\cite{cite_quantum_internet_advances_gyongyosi}. These networks apply quantum mechanics principles of quantum entanglement, teleportation, etc., to transfer data between two remote nodes. Due to quantum channel losses with distances and over time, a multi-hop route with quantum repeaters is prepared to connect remote nodes. Each node has quantum and classical interfaces with its neighbours. Also, nodes generate and transmit entangled pairs, and also store entangled pairs in quantum memory for future usage. At each hop, entanglement swapping is applied which essentially sets up entanglement between two nodes without a direct quantum link via a third node in the middle which is in turn entangled with the other two. This process is repeated to set up long-distance entanglement \cite{cite_qi_link_layer}.
Quantum teleportation is the process of sending a qubit in any arbitrary state to its peer. This process requires a classical channel to convey the results of quantum measurement on the entangled qubits to its receiver and those are used by the latter to retrieve the original qubit.

For large quantum networks, it is necessary to distribute entanglement among remote nodes so that they can transfer data, e.g., via teleportation, based on user requests. Several protocols have been proposed for entanglement generation and distribution, such as, efficient routing to meet end-to-end capacities \cite{cite_rout_ent_for_qi}\cite{cite_end_to_end_capacities_in_qnet}, opportunistic entanglement generation \cite{cite_opportunistic_ent_dist_qi}, connection-oriented and connectionless entanglement distribution \cite{cite_connection_oriented_decoherence_time} \cite{cite_quantum_internet_network_layer_connection_oriented_connectionless}, shortest path based approach \cite{cite_graph_shortest_path_qent}, and average connection time and average largest entanglement cluster size based mechanisms \cite{cite_practical_avg_connnect_time_qent_cluster}. Some of the works have focussed on-demand based entanglement generation  \cite{cite_qent_on_demand_continus}\cite{cite_qi_quantum_storage_overlay}
\cite{cite_qi_deep_reinforcement_maximizes_source_destination_pairs}
\cite{cite_qi_multiple_ent_maximize_number_of_pairs_and_throughput}
\cite{cite_qi_greedy_algo_for_ent_swaps_to_meet_demand}.
The need for proactive entanglement distribution to meet application demands has also been highlighted \cite{cite_qi_quantum_storage_overlay}
\cite{cite_qi_proactive_vs_reactive_ent_distribution}\cite{cite_qi_proactive_ent_historical_count}.

Most of the prior works on quantum networks assume topologies for the perspective of the link layer only and define mechanisms for entanglement generation rate and distribution. However, as observed in the classical internet, the internet hop-count has scaling laws \cite{cite_scaling_law_of_hopcount}\cite{cite_measure_hopcount_gaussian} and the application world wide web shows scale-free properties \cite{cite_scale_free_power_law_www}. Thus, the deployment pattern of the classical internet behaves quite differently from the usage pattern of the same. Drawing parallels from the classical internet, the usage pattern of the quantum networks can be very different from the underlying link network.  Thus, two graphs are likely to emerge. The lower layers graph deal with the reliable entanglement generation and distribution (referred to as \emph{entangled graph}) whereas the upper layers graph emerges out of the user preferences for end-to-end connection (referred to as \emph{connection graph}). In such a setting, it would be interesting to study how the user behaviour and preferences which build the connection graph impact the usage of entanglement distribution in the entangled graph. The question this work tries to answer is, \emph{Do any network centralities emerge in the entangled graph based on the user requested end-to-end connections in the connection graph?} As discussed latter in this paper, the answer is yes and there are certain centralities and emergent patterns in entanglement distribution.

This paper considers three different types of connection graphs (uniform, gaussian and power law) separately to depict the usage pattern of the users where the two end points which need to be entangled are drawn from the same distribution. Uniform distributions are more applicable to small networks where each node may connect to all others. Gaussian distributions are considered because of their theoretical importance. The world wide web application of the internet shows a scale-free distribution \cite{cite_scale_free_power_law_www} whereas the internet hop counts reveal scaling laws \cite{cite_scaling_law_of_hopcount}\cite{cite_measure_hopcount_gaussian}. Hence, a power law is also included in the connection graph. The entangled graph is the same as in \cite{cite_qi_proactive_ent_historical_count} with proactive entanglement distribution which is in agreement with some of the recent works that advocate such a need under different circumstances \cite{cite_qi_quantum_storage_overlay}
\cite{cite_qi_proactive_vs_reactive_ent_distribution}. Under such a two-graph network topology, the centralities of the entangled graph are studied which are driven by the three different user connection request patterns, in the connection graph. For edge centrality, the frequency of the edges used during entanglement generations across the networks is evaluated since quantum entanglements are used peer-to-peer for information transfer. For node centrality, the growth in the degree of nodes in the entangled graph is studied. Results show that the edge centralities follow a power law (of the form $y = Ax^{-B}$) properties whereas the growth in entanglements and degree centralities with the increase in the number of connections show monomolecular distribution (of the form $y = C-De^{-Ex}$). These results can help in quantum resource management in the network and also suggest the choice of quantum technology to be deployed along edges. For example, a reliable entanglement generation with short decoherence times can be deployed across edges with high centralities since they are more likely to be consumed quickly without losses. To the best of the knowledge of the authors, none of the previous work in the literature has studied this aspect.

The rest of the paper is organized as follows. Section \ref{section_literature_survey} surveys some of the relevant and recent works on entanglement distribution and related topics. The system model is described in section \ref{section_system_model}. Results are discussed in section \ref{section_results}. Section \ref{section_conclusion} concludes this along with some possible future directions of research.
% needed in the second column of the first page if using \IEEEpubid
%\IEEEpubidadjcol
\section{Literature Survey}\label{section_literature_survey}
There is significant interest in quantum networks in the research community.
\cite{cite_qi_vision} presents a vision for the quantum internet.
A key part of the quantum internet is entanglement distribution.
A network coding based approach for entanglement distribution has been proposed \cite{cite_net_coding_vs_swapping}.
A set of novel protocols for entanglement generation simultaneously for multiple user pairs with enhanced performance for quantum repeaters have been developed \cite{cite_rout_ent_for_qi}\cite{cite_end_to_end_capacities_in_qnet}.
Application of graph state rather than maximal entanglement of EPR pairs for long-distance routing has been described \cite{cite_q_routing_local_complementation}. \cite{cite_ent_gradient_routing_qi} recommends parallel routing of feasible paths to destinations using the entanglement gradient coefficient to evaluate the shortest distance.
\cite{cite_opportunistic_ent_dist_qi} describes a mechanism to opportunistically distribute entanglements using local minimums of a cost function with quantum memory error and predict evolved entanglement fidelities.
A hybrid entanglement distribution scheme between nodes using entanglement swapping with different quantum devices has been presented in \cite{cite_quantum_swap_hetero_encoding}.
Formulation of the problem to maximize entanglement distribution rate satisfying end-to-end fidelity requirements has been put forward in \cite{cite_multicommodity_flow_qent}.
To reduce the influence of quantum decoherence, \cite{cite_connection_oriented_decoherence_time} proposes a connection-oriented entanglement distribution protocol.
\cite{cite_graph_shortest_path_qent} applies graph based routing mechanism for finding the shortest paths to set up entanglement.
\cite{cite_practical_avg_connnect_time_qent_cluster} uses average connection time and average largest entanglement cluster size to extend entanglement with repeaters within quantum memory decoherence time.
To minimize network latency, \cite{cite_qent_on_demand_continus} applies distributed routing with continuous on-demand entanglement generation. \cite{cite_fidelity_guarantee_qent} proposes an iterative routing algorithm for purification during multiple source-destination EPR  generation with guaranteed fidelity.
\cite{cite_opt_routing_qnet_stochastic} takes into account the parameters, such as, entanglement generation rate, decoherence time, etc., and applies stochastic methods to study optimal routing.

\cite{cite_quantum_internet_network_layer_connection_oriented_connectionless} proposes protocols for connection and connectionless network of quantum routers along with a hybrid combination for the network layer of the protocol stack. The application of quantum paths and their usage in multipartite entanglement distributions has been discussed in  \cite{cite_quantum_paths_for_quantum_internet}. The need to take a global view for the deployment of large-scale quantum internet and the importance of network topology has been highlighted in \cite{cite_qi_network_topology}. A quantum overlay storage network to store EPR pairs as a proactive step to handle subsequent network loads to speed up delays has been proposed in \cite{cite_qi_quantum_storage_overlay}. Using a markov decision process and reinforcement learning, considering short coherence times, link losses and asymmetric links, \cite{cite_qi_high_fidelity_ent_rl_mdp} proposes fast long-distance entanglement distribution with high fidelity. \cite{cite_qi_ent_access_control_determine_tx_rx_without_classical_channel} proposes a protocol to determine the transmitter and receiver to use the EPR pair in multipartite entanglements without using classical channel signaling.
A Sagnac-based orbital angular momentum sorter has been used for multiplexed continuous variable entanglement, which is distributed and measured with the different users in a quasi-complete star network configuration, has been demonstrated in \cite{cite_qi_multiplexed_continous_variable_ent_orbital_angular_momentum}.
\cite{cite_qi_connection_oriented_ent_dist} proposes a connection-oriented entanglement distribution protocol with guaranteed and reliable  EPR generation to reduce the effect of quantum decoherence.
Using maximally entangled (GHZn) states and graph state formalism, \cite{cite_qi_multipartite_ent_GHZn_local_meas_one_qubit_storage} proposes an interesting multipartite entanglement distribution with only local measurements at nodes and with one qubit of memory per user for any arbitrary network topology.
\cite{cite_qi_multpartite_ent_noisy_repeaters_and_memory} proposes an algorithm to generate multipartite entanglement among quantum nodes with noisy quantum repeaters and imperfect quantum memories which is optimal for 3 qubits GHZ state and maximizes both final state fidelity rate of entanglement distribution and final state fidelity.
A deep neural network based routing mechanism to find a path that maximizes the number of source and destination nodes within a time window observing the current state of the network has been proposed in \cite{cite_qi_deep_reinforcement_maximizes_source_destination_pairs}. % No user behaviour
\cite{cite_qi_k_ent_routing_algos} proposes two algorithms, namely, Sequential Multi-path Scheduling Algorithm and Min-Cut-based Multi-path Scheduling Algorithm to perform $k$-entangled routing which connects all $k$ source and destination pairs. % how to decide k which depends on user requests
A mechanism to maximize the number of peer users and also their throughput has been modeled in \cite{cite_qi_multiple_ent_maximize_number_of_pairs_and_throughput}. Using $n$-qubit Greenberger-Horne-Zeilinger (GHZ) measurements,  \cite{cite_n_ent_generation_GHZ_state} proposes a novel method to generate $n$-entanglements each with EPR pairs along $n$ edges at a node.
\cite{cite_qi_high_fidelity_ent_with_photonic_switchboard} describes the distributed generation and automatically prioritized entanglement flows using a quantum router with quantum memories in a photonic switchboard to enable entanglements having high fidelity.
\cite{cite_qi_greedy_algo_for_ent_swaps_to_meet_demand} proposes a greedy algorithm to generate a path for entanglement swapping and maximize the number of satisfied demands concerning start and end times along with the number of entangled pairs.
A mechanism to find the shortest path for entanglement distribution with nested purification of qubits has been proposed in \cite{cite_shortest_path_with_nested_purification}.
Authors present a real-time and pre-established entanglement distribution in \cite{cite_qi_proactive_vs_reactive_ent_distribution}. \cite{cite_qi_hybrid_continuous_discrete_variable_ent} describes a hybrid continuous variable-discrete variable (CV-DV) quantum network which can distribute a large number of entanglements for multi-hop nodes. \cite{cite_qi_proactive_ent_historical_count} proposes a proactive entanglement distribution using the history of actual qubits used along each physical link. The history is an indirect measure of user demands.

From the above survey, it is evident that there is a huge effort in ensuring quick and efficient entanglement distribution. Some of the works have also focussed on meeting user demand for entanglement generation. Most of the proposals are based on a bottom-up approach to meet user demands. The main motivation of this work is to understand how usage patterns can impact entanglement distribution in terms of edge and node centralities. As observed from the above survey, such an effort to understand the network centralities of entanglement distribution has not been attempted. This work is thus relevant as the quantum internet grows in size and its characteristics emerge to its classical counterpart. 
\section{System Model}\label{section_system_model}
Firstly, the basics of quantum communications are discussed and then the model is explained.
\subsection{Quantum Basics}
\subsubsection{Quantum States}
Quantum bits (qubits) are in a superposition of their orthonormal basis states. A single qubit can be represented as $|\phi\rangle =  \alpha|0\rangle + \beta|1\rangle$ where $|\alpha|^2 + |\beta|^2 = 1$, $|0\rangle$ and $|1\rangle$ are the orthonormal basis states. Similarly, a two-qubit can be represented as $|\psi\rangle = \gamma_1|00\rangle + \gamma_2|01\rangle + \gamma_3|11\rangle + \gamma_4|10\rangle$ where $|\gamma_1|^2 + |\gamma_2|^2 + |\gamma_3|^2 + |\gamma_4|^2 = 1$, $|00\rangle$, $|01\rangle$, $|11\rangle$ and $|10\rangle$ are the orthonormal basis states. Measurement of a qubit leads to the collapse of the qubit to one of its basis states. For example, $\psi$ when measured will collapse to one of its four orthonormal states with probability $|\gamma_x|^2$, where $x = 1, 2, 3, 4$.
\subsubsection{Quantum Entanglement}
A two-qubit system can be in a special state which cannot be expressed as a tensor product of two other states. Lets consider a qubit $|\psi_1\rangle = \delta_1|01\rangle + \delta_2|11\rangle$. This can be represented as $|\psi_1\rangle = (\delta_1|0\rangle + \delta_2|1\rangle) \otimes |1\rangle$ where $\otimes$ is the tensor product. However, a special two-qubit of the form $|\psi_2\rangle =\delta_1|00\rangle + \delta_2|11\rangle$ cannot be represented as a tensor product to two states as $|\psi_1\rangle$. These special states are called entangled states. There are four maximally entangled states also known as Bell pairs or EPR pairs, namely, $\frac{|00\rangle + |11\rangle}{\sqrt{2}}$, $\frac{|00\rangle - |11\rangle}{\sqrt{2}}$, $\frac{|01\rangle + |10\rangle}{\sqrt{2}}$ and  $\frac{|01\rangle - |10\rangle}{\sqrt{2}}$. If the first qubit of the EPR pair is given to one node and the other to the second, and a measurement is performed on the EPR pair then their states are highly deterministic. For example, $\frac{|01\rangle + |10\rangle}{\sqrt{2}}$ will collapse with the first node having a value $0$ and the other having a value $1$, and vice versa when measured.
\subsubsection{Quantum Entanglement Swapping}
If a node $A$ is entangled to $B$  and $B$ in turn is entangled to $C$ then $A$ and $C$ can be entangled with the measurement of the two individual entanglements. This is known as Quantum Entanglement Swapping. This process can be repeated over multiple hops to create a long-distance entanglement.
\subsubsection{Quantum Teleportation}
In a nutshell, quantum teleportation is a method of transferring any arbitrary qubit $\Psi$ from one source to another using a Bell pair shared between them. It requires a classical channel to communicate the results of the measurement to the receiver. Using the result, the receiver applies linear transforms to recreate $\Psi$.
\subsection{Model Description}
Let there be $N$ nodes in a quantum network. There are $E$ edges in the network indexed with $i$. A node $j$ can have uniform random number of edges $X^{(p)}_j$ in the range $1$ to $\lfloor\alpha N\rfloor$, $0 < \alpha \le 1$, with other nodes. These $X^{(p)}_j$ are designated as physical edges which can be used to set up direct entanglements between the connected nodes.

These entangled physical edges can help in setting up entanglement between nodes that are not physically connected using quantum entanglement swapping. For example, if node A is entangled with B and B in turn is entangled with C then applying entanglement swapping  A and C can get entangled. Such edges which connect two remote entangled nodes set up using entanglement swapping are designated as virtual edges $X^{(v)}_j$. Thus, the total number of edges that connect node $j$ to its entangled neighbours (physical and virtual) is denoted by $X_j$. Thus,
\begin{equation}
X_j = X^{(p)}_j + X^{(v)}_j
\end{equation}

This process of entanglement generation either directly through a physical link or indirectly through entanglement swapping can provide valuable insights (e.g., the centrality of edges) on how each physical edge $e^{(p)}_i$ or a virtual $e^{(v)}_i$ edge is involved in entanglement generation for a large quantum network when entanglements are needed between two remote nodes based on user connection setup requests $k$. Traditionally, edge centrality measures are based on the shortest paths between nodes. Since, quantum entanglements are used on a peer-to-peer basis (when measurements are performed) usage frequencies ($f_{e^{(p)}_i}$ or $f_{e^{(v)}_i}$) of links during the entanglement generation and distributions are used as a measure of edge centrality in this paper. Generally, $f_{e_i}$ denotes the usage frequency of any edge either physical or virtual in an entanglement setup. The higher the usage of a link the larger is its edge centrality.

As mentioned above, an edge (physical or virtual) is created in the graph of the quantum network when the two nodes share an entanglement. In the process of edges getting added, the degree centrality (the degree of the node) of the nodes also increases. Thus, degree centralities can also provide a key insight into quantum resource allocation. The degree centrality of node $j$ is designated with $d_j$.

In \cite{cite_qi_proactive_ent_historical_count}, authors proposed a mechanism for proactive entanglement distribution among some nodes quicker end-to-end connection setup in large quantum networks. The proactive entanglement distributions are based on the historical count of qubits transmitted along the edges. 
The process described in \cite{cite_qi_proactive_ent_historical_count} involves the following steps. Each physical edge has a certain historical count of qubits transferred over it. To set up proactive entanglements, some of the nodes are uniformly randomly selected. Each node first calculates the mean of historical counts with its neighbours. It then calculates the squared differences between the mean and the corresponding historical counts of its neighbours. The node with the minimum difference is chosen for the entanglement setup. This process is performed by all randomly selected nodes to set up entanglements.
After these entanglements are set up, entanglement swaps are applied wherever applicable to set up further entanglements. These entanglements are set up proactively and saved in quantum memory to help in speeding up the end-to-end connection requested by the user at a later point in time. Based on the user connection setup request, the end-to-end entanglement is setup which is later used by teleportation for data transfer. A classical channel is needed for teleportation and also for routing the connection setup message \cite{cite_qi_proactive_ent_historical_count}.

When a large number of connections are set up in the network, two graphs emerge. One is for the entanglements (entangled graph) and the other is for end-to-end user connection requests (connection graph). The paper aims to study the emergent properties of the entanglement graph as a large number of connections are set up with different user request patterns. Three scenarios are considered where the two end nodes for connection setups are drawn from the same uniform, gaussian and power law distribution. Uniform distribution may apply to a small quantum network. Considering today's world wide web follows a power law, the application of the same on a quantum network has practical importance. Gaussian distribution is studied for its theoretical significance.
\begin{figure}[ht]
\centering
\includegraphics[width=\columnwidth]{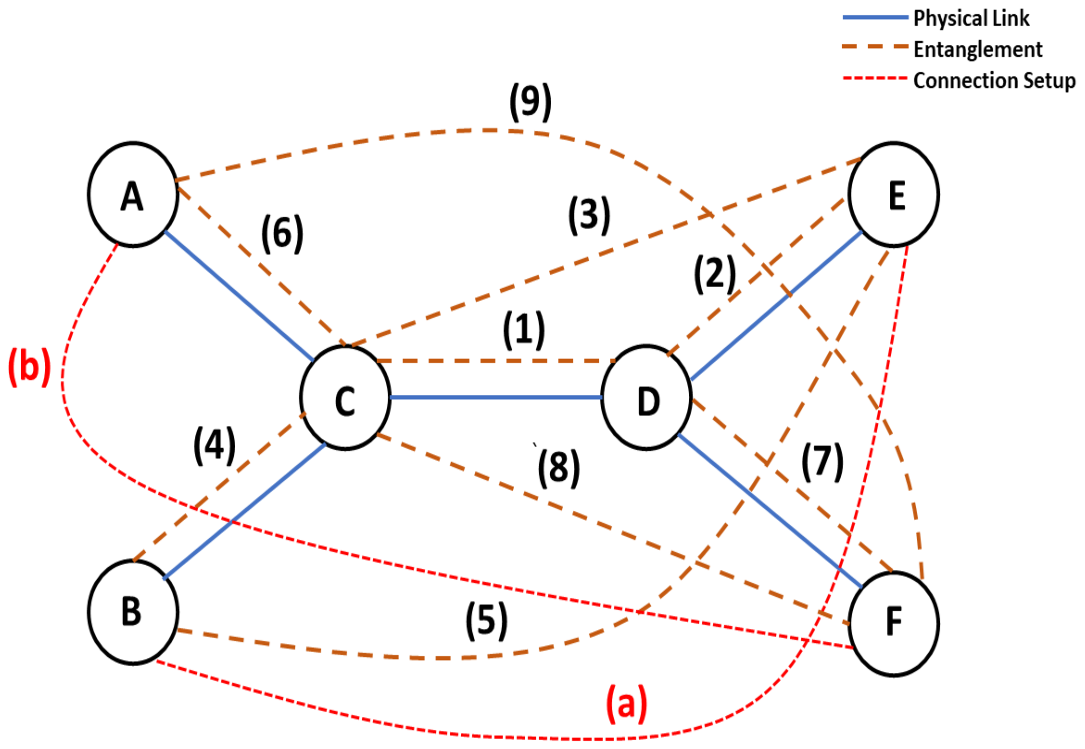}
\caption{Example Quantum Network}
\label{plot_Qnet_ink}
\end{figure}

Fig. \ref{plot_Qnet_ink} shows an example quantum network. All blue lines show the physical quantum links, brown dashed lines are the entanglements set up either directly or through entanglement swaps and red lines with small dashes are the connection requests from users. For example, entanglements (1) and (2) setups are based on HCs proactively as explained in \cite{cite_qi_proactive_ent_historical_count}. (3) is set up with an entanglement swap proactively. The usage frequencies of (1) and (2),  namely, $f_{e^{(p)}_1}$ and $f_{e^{(p)}_2}$ respectively, increase by one to create (3). Let's assume there is a user connection request (a) for B and E. Now, B does not have any entanglement, it sets up with C and increases $f_{e^{(p)}_4}$. To set up an entanglement between B and E, an entanglement swap with (3) and (4) will create entanglement (5) and complete connection (a). This will increase the usage frequencies $f_{e^{(v)}_3}$ and $f_{e^{(p)}_4}$. Let's consider another user connection request (b) between A and F. Node A sets up entanglement C, i.e., (6). When the connection request reaches D, it sets up entanglement with F, i.e., (7). Proactively, entanglement (8) is set up. This process futher increases $f_{e^{(p)}_1}$ and $f_{e^{(p)}_7}$. Finally, a Q-SWAP with (6) and (8) creates (9) and increase $f_{e^{(p)}_6}$, $f_{e^{(p)}_8}$ and $f_{e^{(p)}_9}$ respectively.

This process of connection setup also leads to growth in the degree centralities of the nodes in terms of quantum entanglements. The final degree centralities, i.e., $d_j$s, namely, $d_A$, $d_B$, $d_C$, $d_D$, $d_E$ and $d_F$, are 2, 2, 5, 3, 3 and 3 respectively.

The emergent behaviours of the following parameters are studied from the edge centrality perspective.
1) The behaviour of usage frequencies of edges during entanglement distribution, i.e., $sf_{e_i}$s'. Also, it would be interesting to see the usage patterns of physical (i.e., $f_{e^{(p)}_i}$s')  and virtual edges (i.e., $f_{e^{(v)}_i}$s') separately for three distributions of user connection setup requests (uniform, gaussian and power law). This will provide an understanding of the centralities of the edges where quantum resources are likely to be more consumed.
2) The behaviour of the above frequencies with the increase in the number of connections for the three distributions is also interesting since it explains whether the observed behaviour scale with the number of connections.
3) For the gaussian distribution, the behaviour of the above frequencies with the increase in standard deviations is also studied.
4) For power law distribution, the same behaviour is studied with decreasing exponent
5) Also, it would be beneficial to compare the behaviour of these usage frequencies for all three distributions.
6) The behaviour of entanglement build up with increasing connections for the three distributions will indicate the dynamics of the overall network.
7) As a use case, it would be valuable to find out how the edges with maximum usage, i.e., $max(f_{e_i})$, $max(f_{e^{(p)}_i})$ and $max(f_{e^{(p)}_i})$ behave with the increasing number of connections for all the three distributions.

Some of the behaviours of interest from node perspective are as follows.
1) It would be of interest how the degree of a node $d_j$ evolves as a function of the number of connections $k$, i.e.,  $d_j(k)$ for the three connection setup distributions.
2) Also, it is important to find out the cummulative growth in the degree of nodes, i.e., $\sum_{k=1}^{M} [d_j(k+1) - d_j(k)]$, where $M$ is the maximum number of connections, for the three user connection setup patterns.

The model assumes the following points including those from \cite{cite_qi_proactive_ent_historical_count}.
1) Once entanglement between two is set up it is perpetually used.
2) Reliable generation of qubits is ensured between any two entangled nodes.
3) Only one connection setup attempt is made between two remote nodes. Unlike \cite{cite_qi_proactive_ent_historical_count}, multiple retries are not allowed.
4) Each node has classical and quantum interfaces.
5) Each node can generate entanglement directly over the physical quantum link or virtual using an entanglement swap.
6) Each node saves entangled qubits in stable quantum memory for data transfer. The historical counts are stored in classical memory.
7) All control procedures, such as, connection setup procedures are performed over the classical interface and data transfer through teleportation over entangled edges (with support from the classical interface).
\section{Results and Discussion}\label{section_results}
This section presents the simulation results based on the above system model. The simulation model is implemented in R language. The parameters used in the simulation are listed in table \ref{table_simulation_parameters}.
\begin{table}[ht]
  \caption{Simulation Parameters}
  \centering
  \begin{tabular}{|p{1.5cm}|p{1.5cm}|p{4.5cm}|}
  \hline
  Parameter & Value & Description\\  [0.5ex]
  \hline
  $N$ & $100$ & Maximum number of nodes\\
  \hline
  $M$ & $10^3$, $10^4$, $10^5$ & Maximum number of connections\\
  \hline
  $\alpha_i$ & $25\%$ & Maximum percentage of physical quantum links of $i^{th}$ node\\
  \hline
  $\mu$ & $50$ & mean for gaussian distribution\\
  \hline
  $\sigma$ & $20$ & standard distribution for gaussian distribution\\
  \hline
  $B$ & $-0.75$ & power law exponent\\
  \hline
  $A$ & $10^2$, $10^3$, $10^4$ & power law coefficients for $10^3$, $10^4$, $10^5$ connections respectively\\
  \hline
  \end{tabular}
  \label{table_simulation_parameters}
\end{table}
\subsection{Edge properties}
\subsubsection{Connection setup endpoints drawn from uniform distribution}
For connection setup, any two nodes from the quantum network of 100 nodes are drawn from uniform distribution and a hundred thousand connection setup requests are made.
%\subsubsection{Entangled edges - physical and virtual}

\begin{figure}[ht]
\centering
\includegraphics[width=\columnwidth]{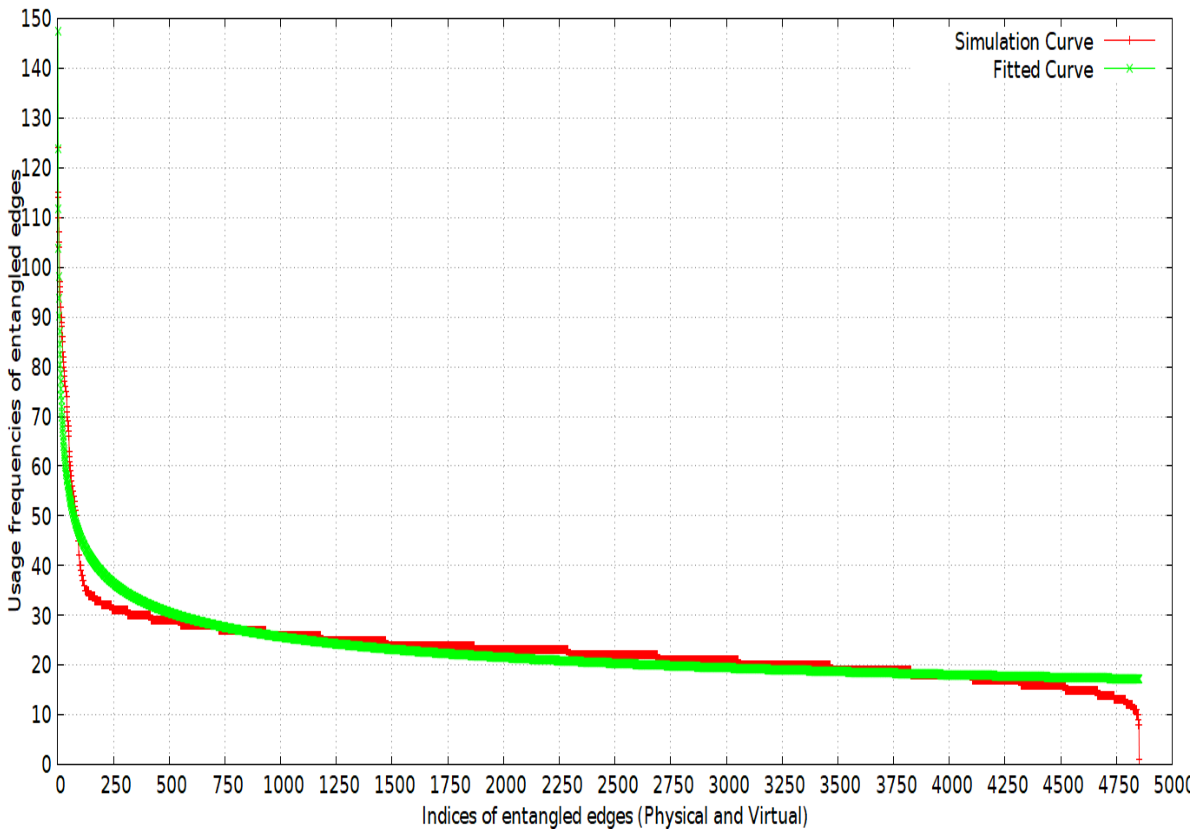}
\caption{Connection Setup - Uniform Distribution, physical and virtual entanglements, Curve fit $f_{e_{i}} = 147.4696i^{-0.253105}$ where $i$ is the edge}
\label{plot_uniform_all_ent_edges_freq_vs_edge_indices_sorted_curve_fit_ink}
\end{figure}

\begin{figure}[ht]
\centering
\includegraphics[width=\columnwidth]{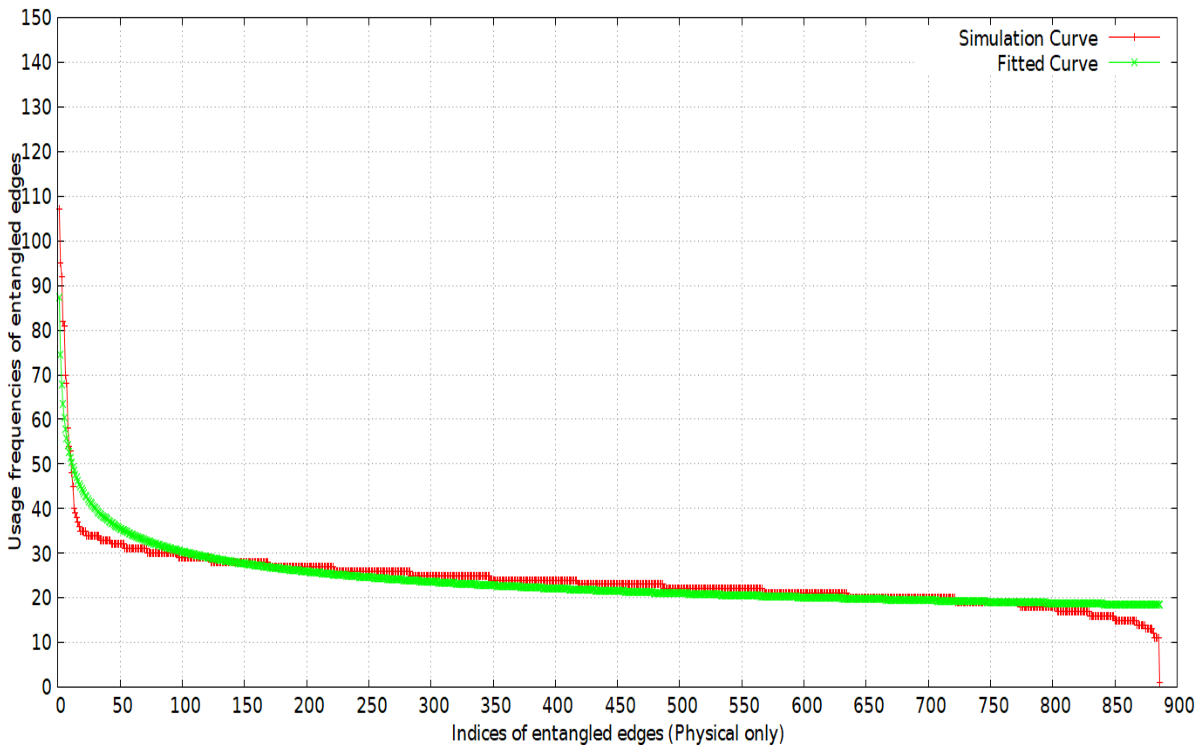}
\caption{Connection Setup - Uniform Distribution, physical entanglements only, Curve fit $f_{e^{(p)}_{i}} = 87.23075i^{-0.2292402}$ where $i$ is the physical edge}
\label{plot_uniform_phy_ent_edges_freq_vs_edge_indices_sorted_curve_fit_ink}
\end{figure}

\begin{figure}[ht]
\centering
\includegraphics[width=\columnwidth]{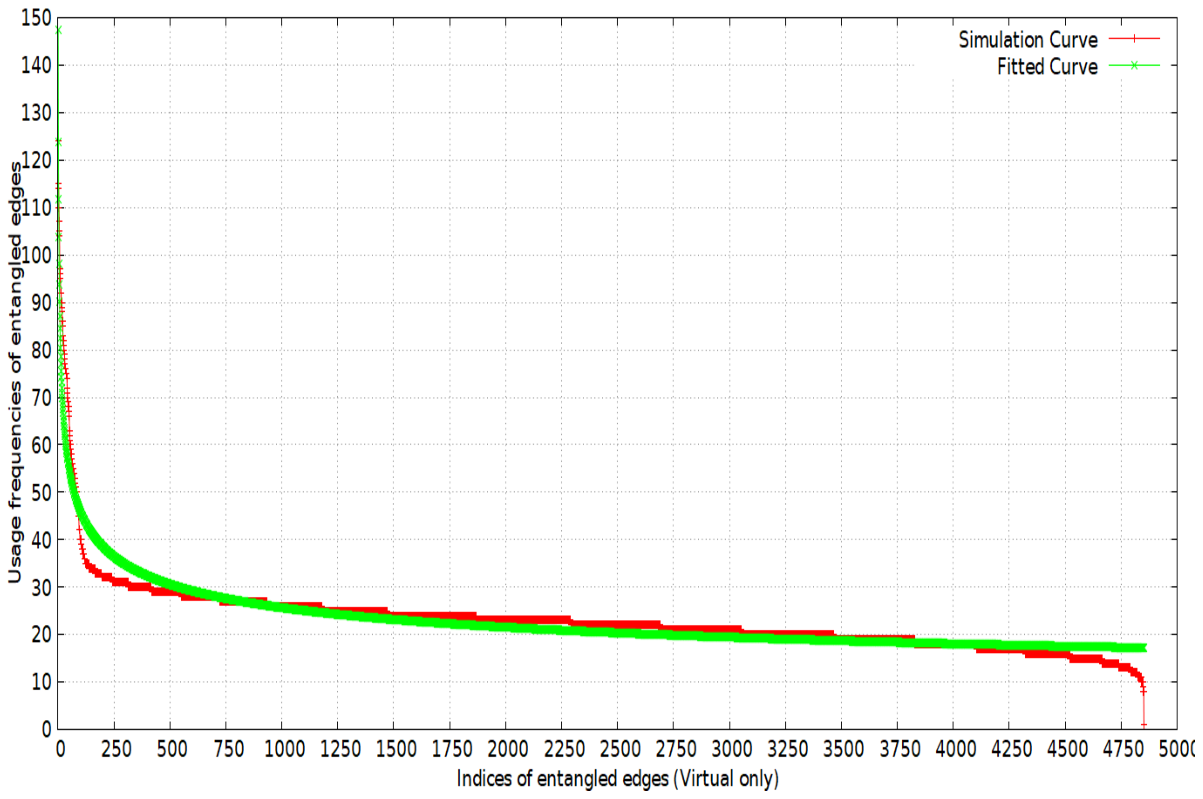}
\caption{Connection Setup - Uniform Distribution, virtual entanglements only, Curve fit $f_{e^{(v)}_{i}} = 147.4696i^{-0.253105}$ where $i$ is the virtual edge}
\label{plot_uniform_virtual_ent_edges_freq_vs_edge_indices_sorted_curve_fit_ink}
\end{figure}

Fig. \ref{plot_uniform_all_ent_edges_freq_vs_edge_indices_sorted_curve_fit_ink} shows the usage of all entangled edges, physical as well as virtual. The indices of the edges (i.e., $i$ of $e^{(p)}_i$ or $e^{(v)}_i$) are shown along \emph{x}-axis whereas the frequencies of usage of entangled edges ($f_{e^{(v)}_i}$ or $f_{e^{(v)}_i}$) used to setup further entanglement is shown along \emph{y}-axis. The result from the simulation is shown with the red line which exhibits a power law behaviour. This is confirmed by the fitted power law curve in green with coefficient $147.4696$ and exponent $-0.253105$. This result shows that even though the nodes for connection setup are drawn from uniform distribution some of the entangled edges are more frequently used than others. More quantum resources need to be assigned to these edges. Also, qubits with technologies with shorter decoherence time but with more reliability may be assigned along these edges since they are more likely to be consumed before they lose fidelity. For low usage edges, qubits with technologies with longer decoherence time may be applied.
%\subsubsection{Entangled edges - physical}
Fig. \ref{plot_uniform_phy_ent_edges_freq_vs_edge_indices_sorted_curve_fit_ink} shows the behaviour for the physical entangled edges only. This also depicts a similar power law pattern as Fig .\ref{plot_uniform_all_ent_edges_freq_vs_edge_indices_sorted_curve_fit_ink} with coefficient $87.23075$ and exponent $-0.2292402$.
%\subsubsection{Entangled edges - virtual}
Fig. \ref{plot_uniform_virtual_ent_edges_freq_vs_edge_indices_sorted_curve_fit_ink} shows the behaviour for the virtual entangled edges only. A similar power law pattern is observed with coefficient $147.4696$ and exponent $-0.253105$.
%\subsubsection{Entangled edges vs number of connections}

\begin{figure}[ht]
\centering
\includegraphics[width=\columnwidth]{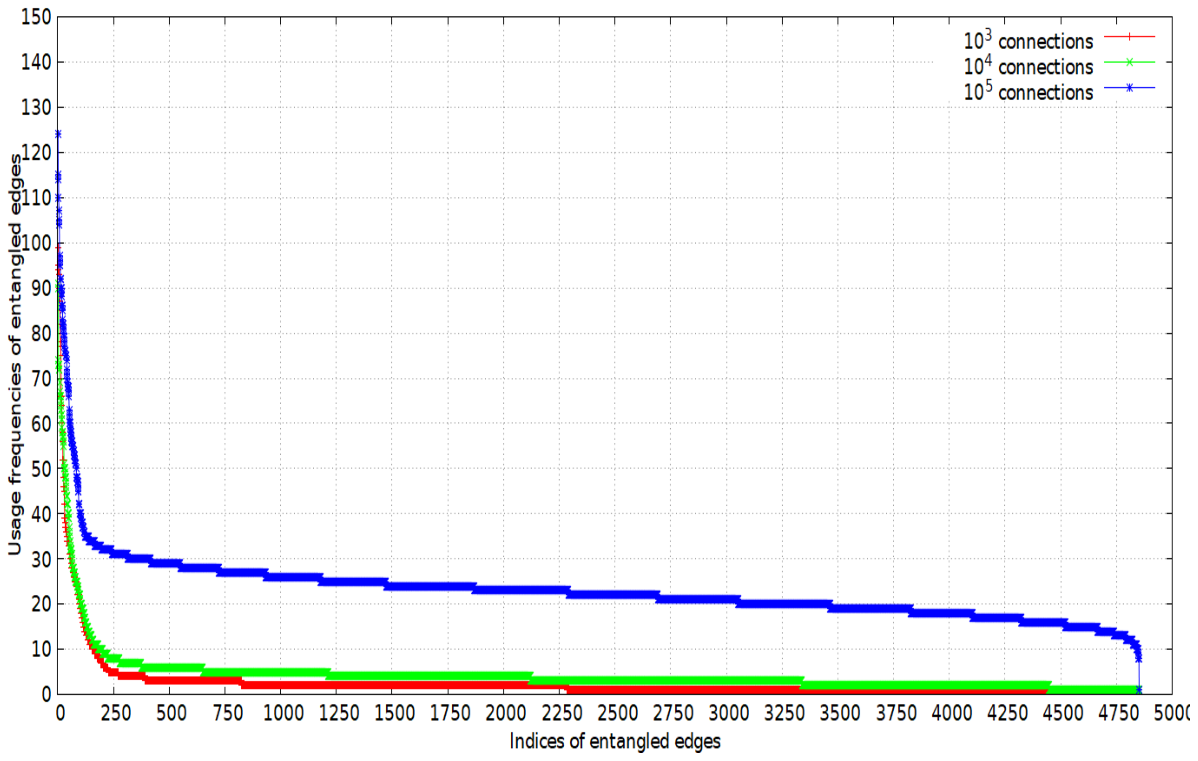}
\caption{Entangled edges usage with the increasing number of connections}
\label{plot_uniform_all_ent_edges_freq_vs_edge_indices_vs_connections_ink}
\end{figure}

\begin{figure}[ht]
\centering
\includegraphics[width=\columnwidth]{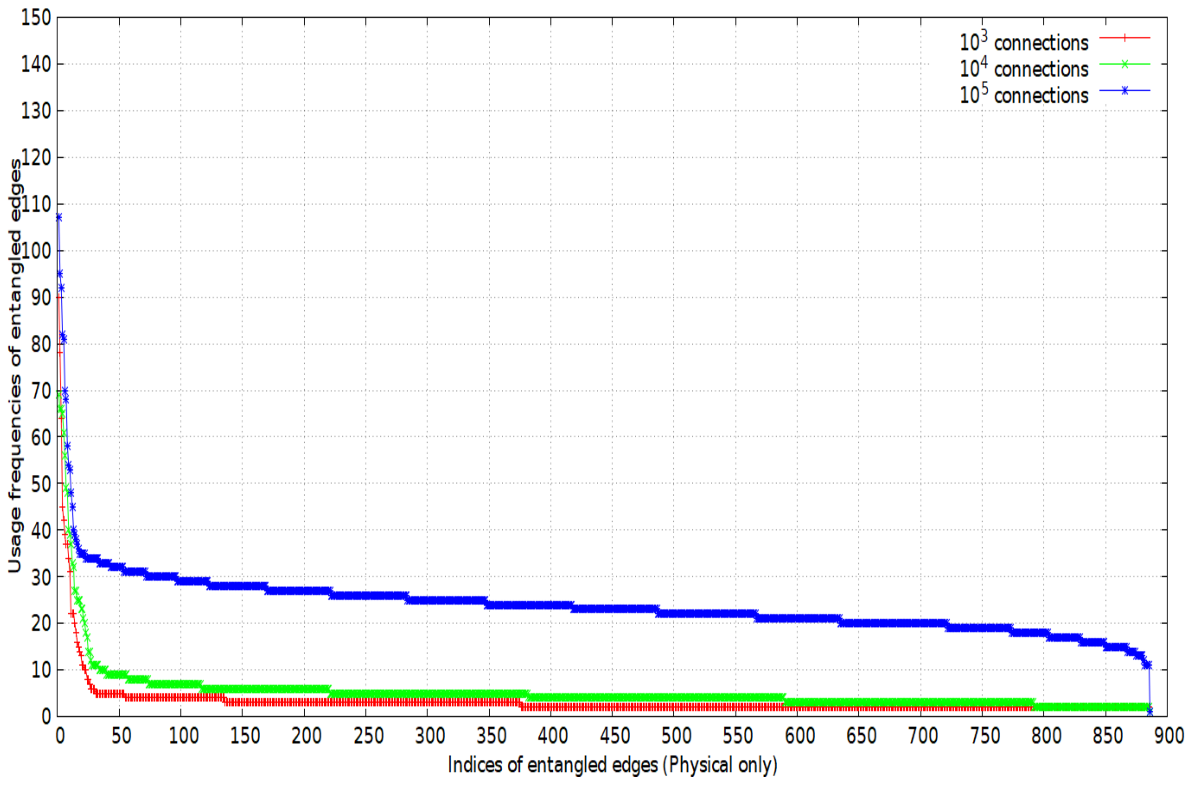}
\caption{Entangled edges (Physical only) usage with the increasing number of connections}
\label{plot_uniform_phy_ent_edges_freq_vs_edge_indices_vs_connections_ink}
\end{figure}

\begin{figure}[ht]
\centering
\includegraphics[width=\columnwidth]{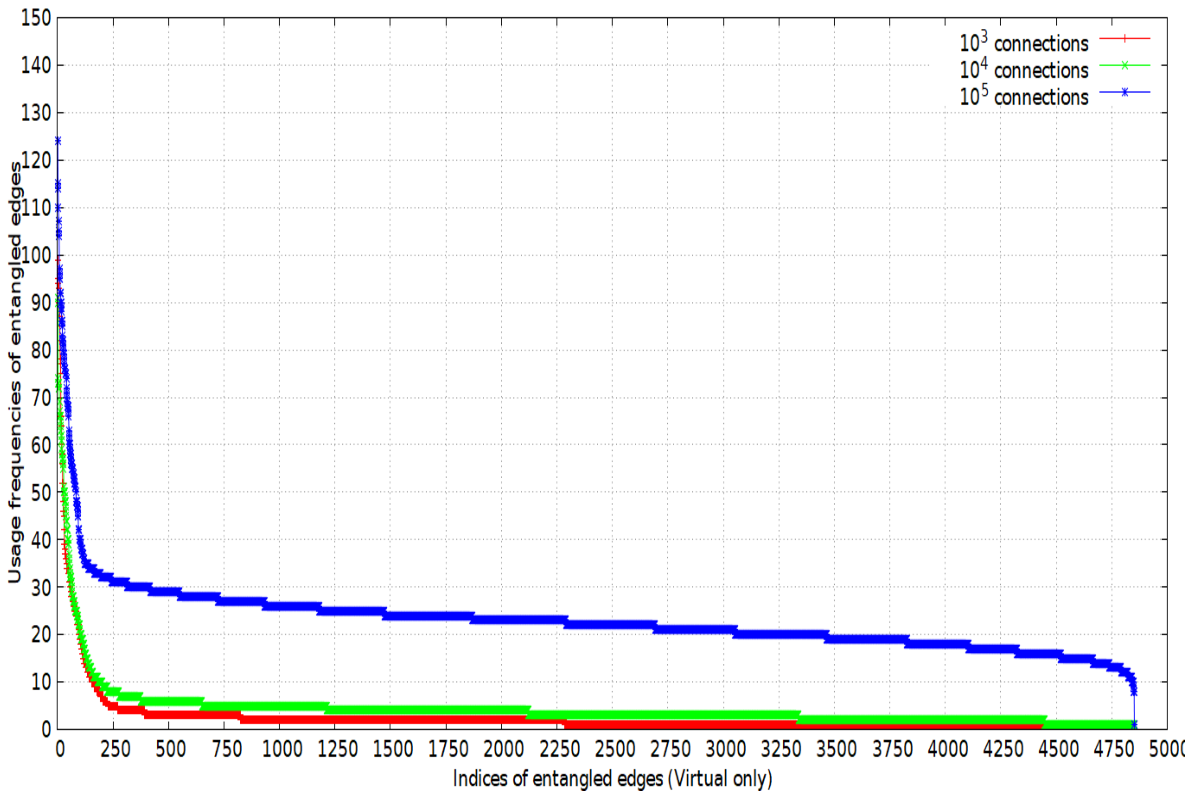}
\caption{Entangled edges (Virtual only) usage with the increasing number of connections}
\label{plot_uniform_virtual_ent_edges_freq_vs_edge_indices_vs_connections_ink}
\end{figure}

The behaviour of usage frequencies of entangled edges (both virtual and physical) with the increasing number of connection requests is shown in Fig. \ref{plot_uniform_all_ent_edges_freq_vs_edge_indices_vs_connections_ink}. The indices of the edges are shown along \emph{x}-axis and the usage frequencies are along \emph{y-axis}. There are a couple of observations. Firstly, as the number of connections increases the usage frequencies of entangled edges also grow. Secondly, roughly the same first 100 edges show exponentially higher usages with an increasing number of connection setup requests. Fig. \ref{plot_uniform_phy_ent_edges_freq_vs_edge_indices_vs_connections_ink} and Fig. \ref{plot_uniform_virtual_ent_edges_freq_vs_edge_indices_vs_connections_ink} show the behaviour of physical and virtual edges respectively with the increasing number of connections. Both of them show similar behaviours with virtual edges showing slightly higher usages than the physical ones.
\subsubsection{Connection setup end points drawn from gaussian distribution}
\begin{figure}[ht]
\centering
\includegraphics[width=\columnwidth]{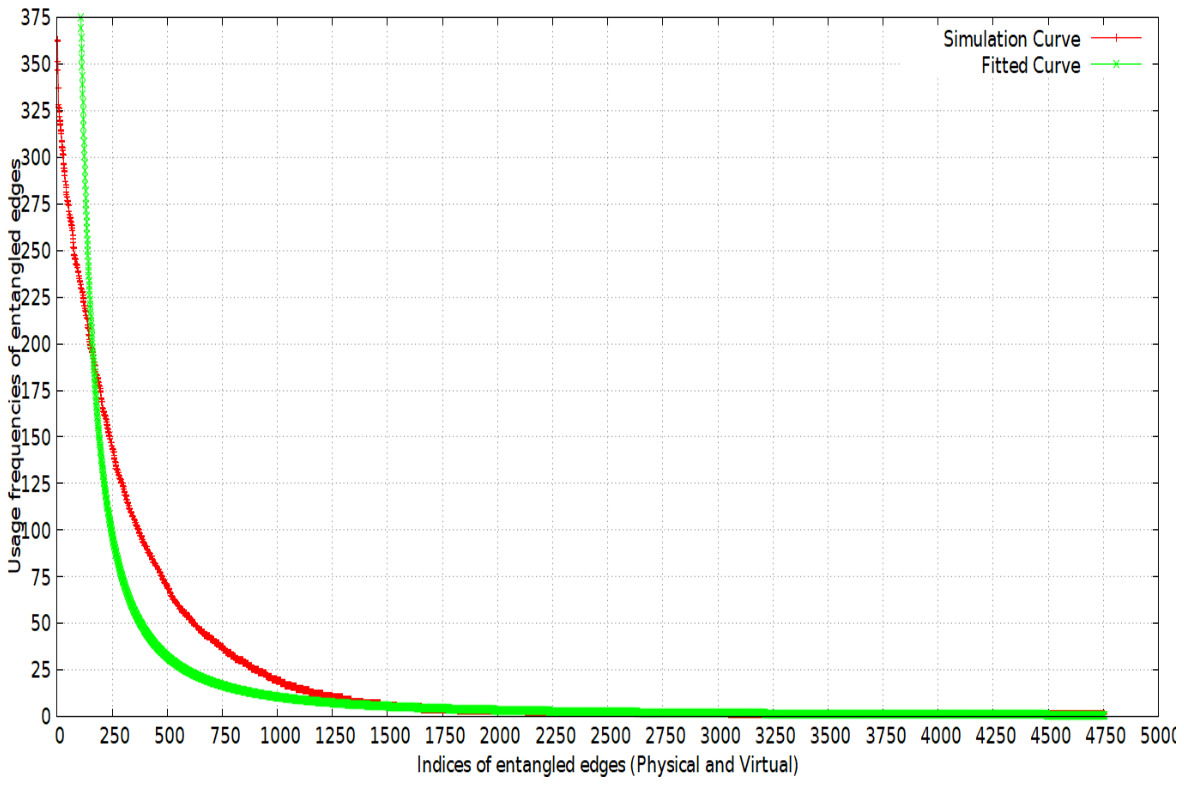}
\caption{Connection Setup - Gaussian Distribution,  physical and virtual entanglements, $f_{e_{i}} = 684876.1i^{-1.604152}$ where $i$ is the edge}
\label{plot_normal_all_ent_edges_freq_vs_edge_indices_sorted_curve_fit_ink}
\end{figure}

\begin{figure}[ht]
\centering
\includegraphics[width=\columnwidth]{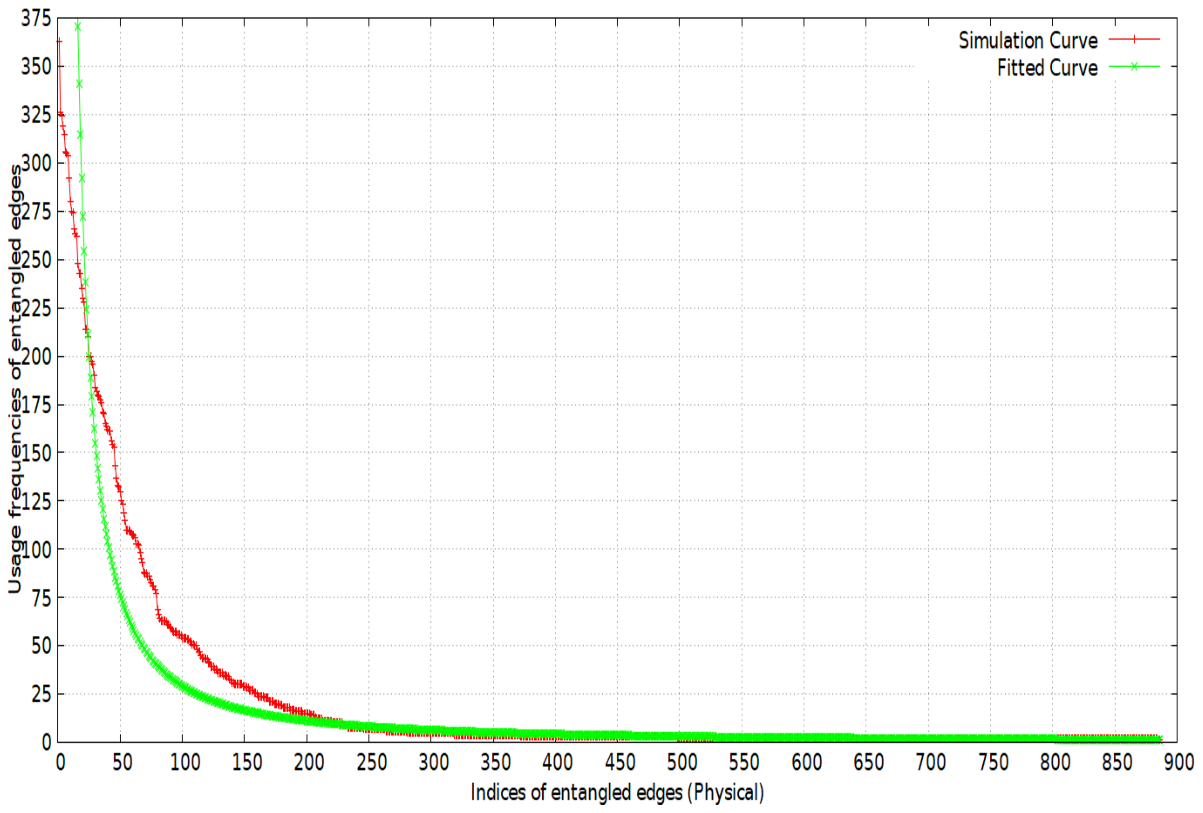}
\caption{Connection Setup - Gaussian Distribution,  physical entanglements only, $f_{e^{(p)}_{i}} = 17348.56i^{-1.387009}$ where $i$ is the physical edge}
\label{plot_normal_phy_ent_edges_freq_vs_edge_indices_sorted_curve_fit_ink}
\end{figure}

\begin{figure}[ht]
\centering
\includegraphics[width=\columnwidth]{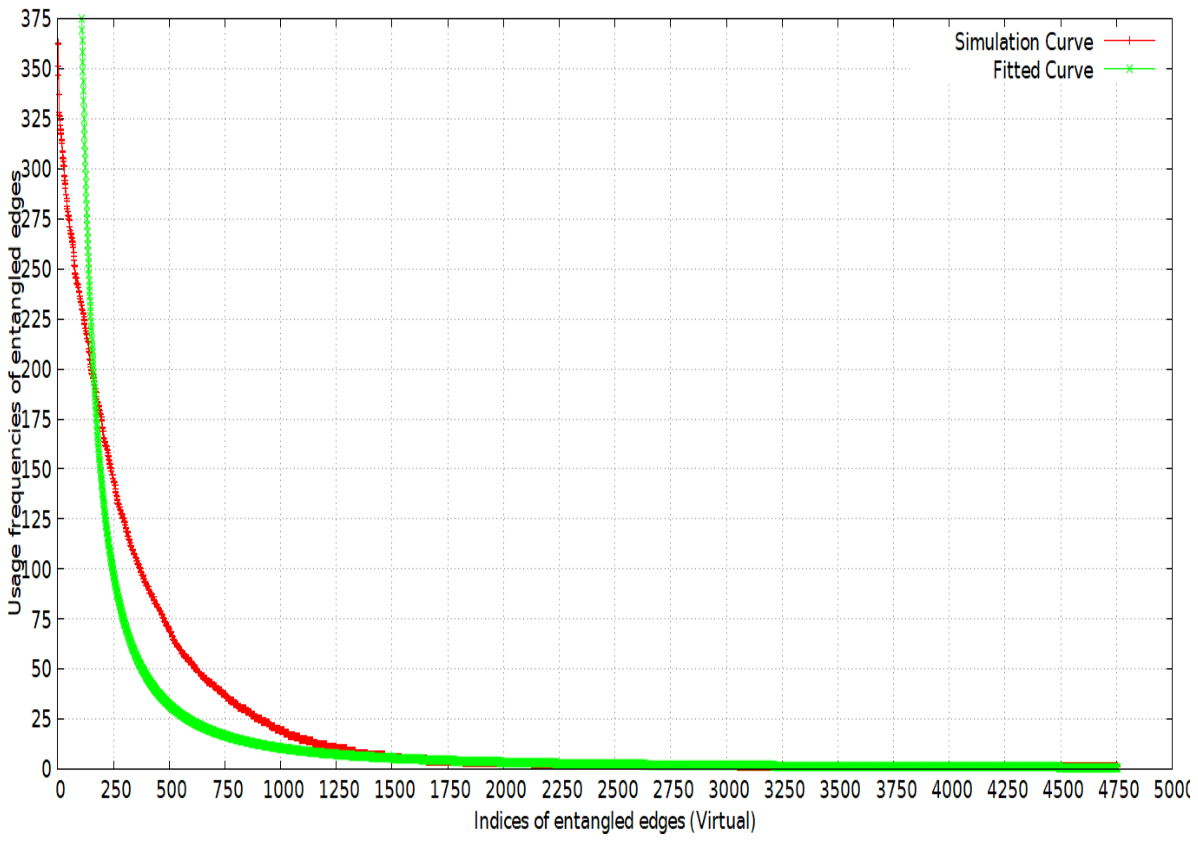}
\caption{Connection Setup - Gaussian Distribution,  physical entanglements only, curve fit $f_{e^{(p)}_{i}} = 684876.1i^{-1.604152}$ where $i$ is the virtual edge}
\label{plot_normal_virtual_ent_edges_freq_vs_edge_indices_sorted_curve_fit_ink}
\end{figure}

\begin{figure}[ht]
\centering
\includegraphics[width=\columnwidth]{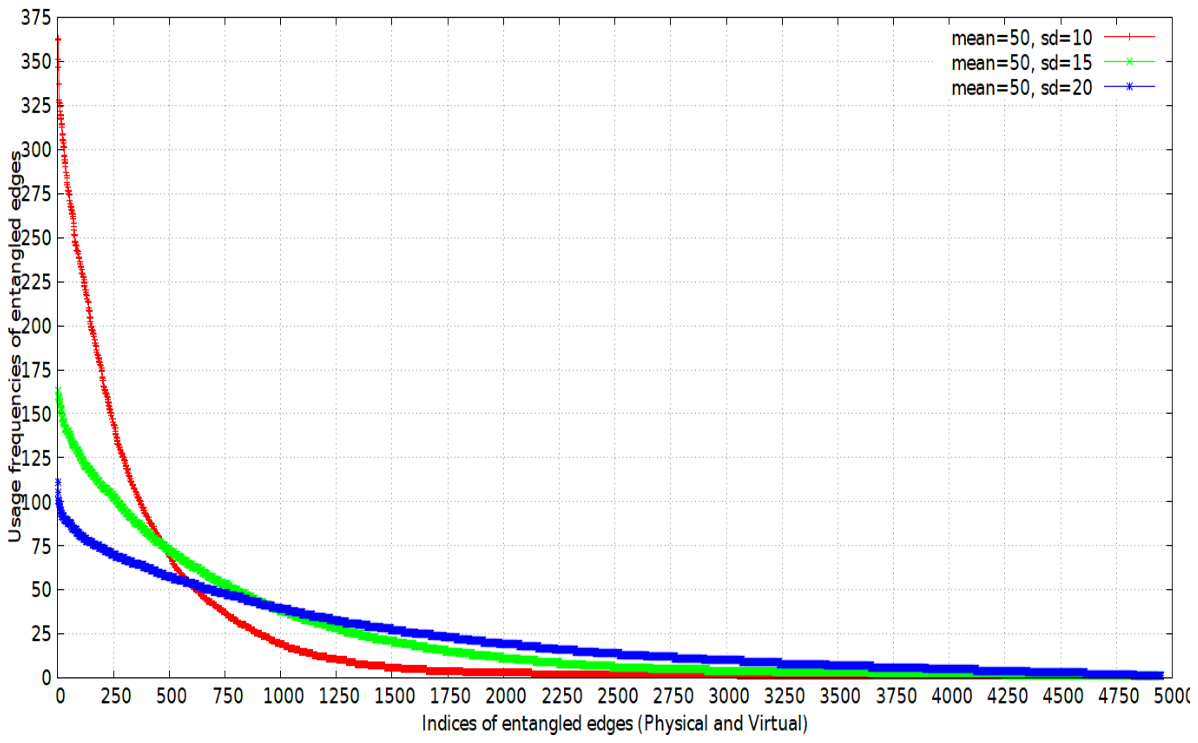}
\caption{Entangled edges (Physical and Virtual) usage with increasing standard deviation and same mean}
\label{plot_normal_increasing_sd_all_ent_edges_freq_vs_edge_indices_sorted_ink}
\end{figure}

\begin{figure}[ht]
\centering
\includegraphics[width=\columnwidth]{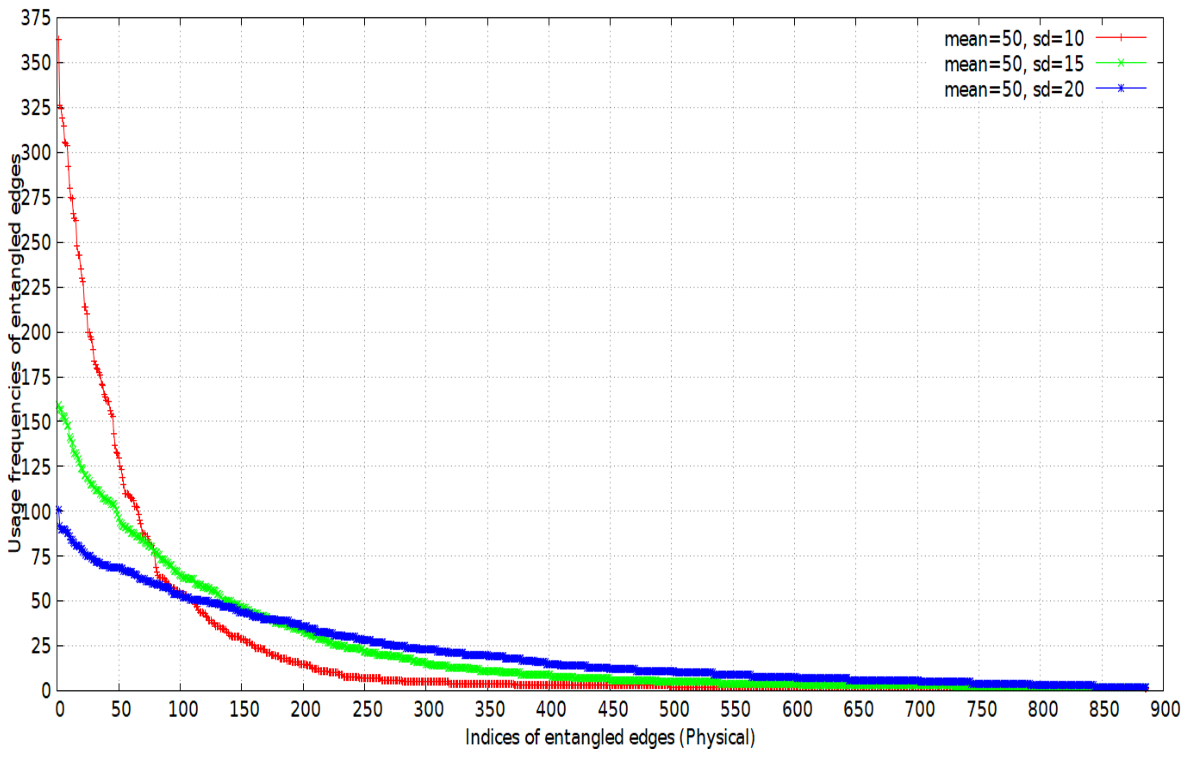}
\caption{Entangled edges (Physical) usage with increasing standard deviation and same mean}
\label{plot_normal_increasing_sd_phy_ent_edges_freq_vs_edge_indices_sorted_ink}
\end{figure}

\begin{figure}[ht]
\centering
\includegraphics[width=\columnwidth]{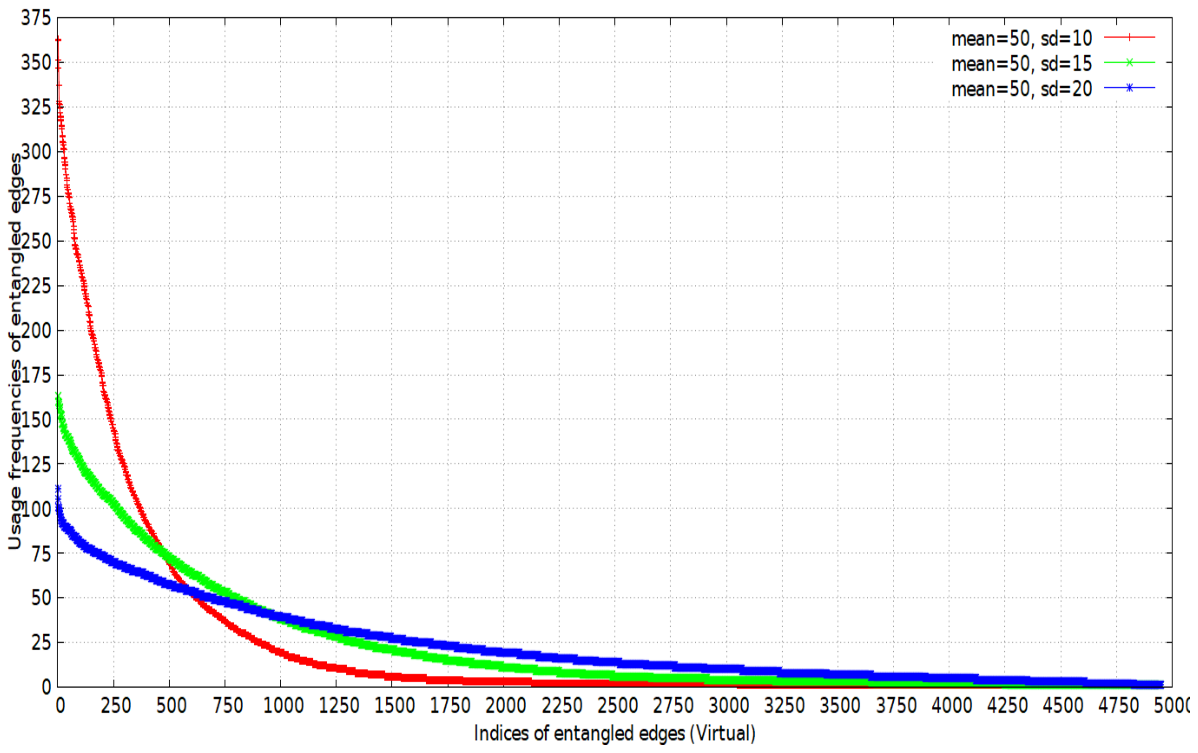}
\caption{Entangled edges (Virtual) usage with increasing standard deviation and same mean}
\label{plot_normal_increasing_sd_virtual_ent_edges_freq_vs_edge_indices_sorted_ink}
\end{figure}

The behaviour of usage frequencies of entangled edges when the nodes for connection setup requests are selected from a gaussian distribution with a mean of 50 and standard deviation of 10 is shown in Fig. \ref{plot_normal_all_ent_edges_freq_vs_edge_indices_sorted_curve_fit_ink}. The indices of the edges are shown along \emph{x}-axis and the usage frequencies are along \emph{y}-axis. The simulation curve in red shows a similar increasing behaviour as uniform distribution. Although, a power law curve fit does not show a good match as before, still it is not that bad either. Fig. \ref{plot_normal_phy_ent_edges_freq_vs_edge_indices_sorted_curve_fit_ink} and Fig. \ref{plot_normal_virtual_ent_edges_freq_vs_edge_indices_sorted_curve_fit_ink} show similar behaviours for physical and virtual entangled edges respectively.
% increasing standard deviation
With the increase in standard deviation keeping the mean same for the guassian distribution there is a decrease in the centrality of the usage of edges as shown in Fig. \ref{plot_normal_increasing_sd_all_ent_edges_freq_vs_edge_indices_sorted_ink}. The physical entangled and virtual entangled edges also show similar behaviour in Fig. \ref{plot_normal_increasing_sd_phy_ent_edges_freq_vs_edge_indices_sorted_ink} and Fig. \ref{plot_normal_increasing_sd_virtual_ent_edges_freq_vs_edge_indices_sorted_ink} respectively.
\subsubsection{Connection setup end points drawn from power law distribution}
\begin{figure}[ht]
\centering
\includegraphics[width=\columnwidth]{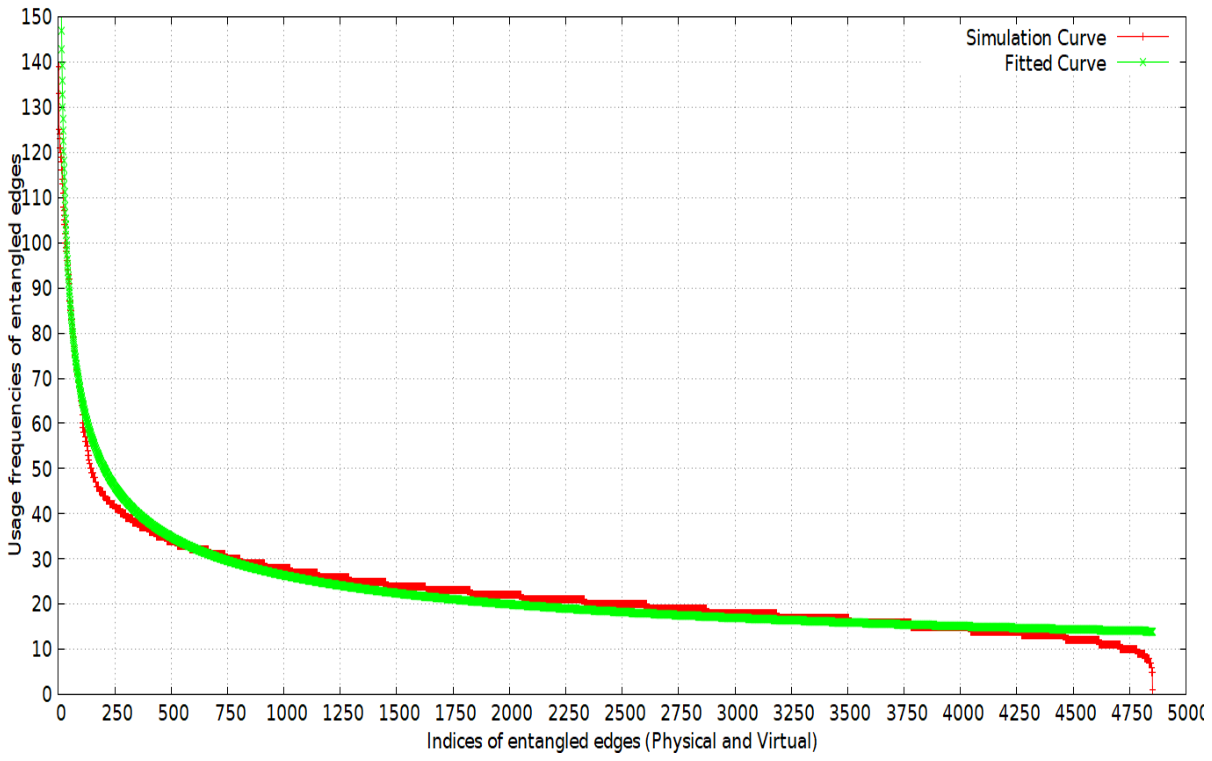}
\caption{Connection Setup - Power Law Distribution, physical and virtual entanglements, curve fit $f_{e_{i}} = 424.8808i^{-0.4023611}$ where $i$ is the edge}
\label{plot_powerlaw_all_ent_edges_freq_vs_edge_indices_sorted_curve_fit_ink}
\end{figure}

\begin{figure}[ht]
\centering
\includegraphics[width=\columnwidth]{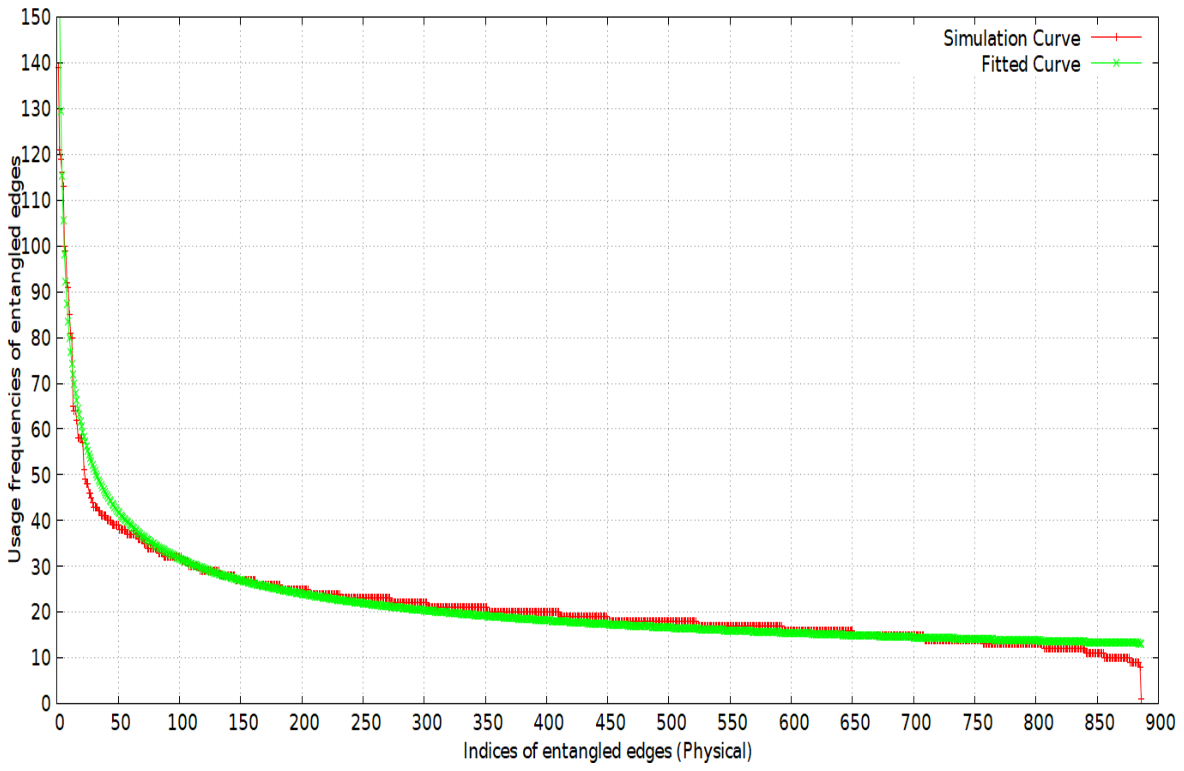}
\caption{Connection Setup - Power Law Distribution, physical entanglements only, curve fit $f_{e^{(p)}_{i}} = 201.4059i^{-0.4014564}$ where $i$ is the physical edge}
\label{plot_powerlaw_phy_ent_edges_freq_vs_edge_indices_sorted_curve_fit_ink}
\end{figure}

\begin{figure}[ht]
\centering
\includegraphics[width=\columnwidth]{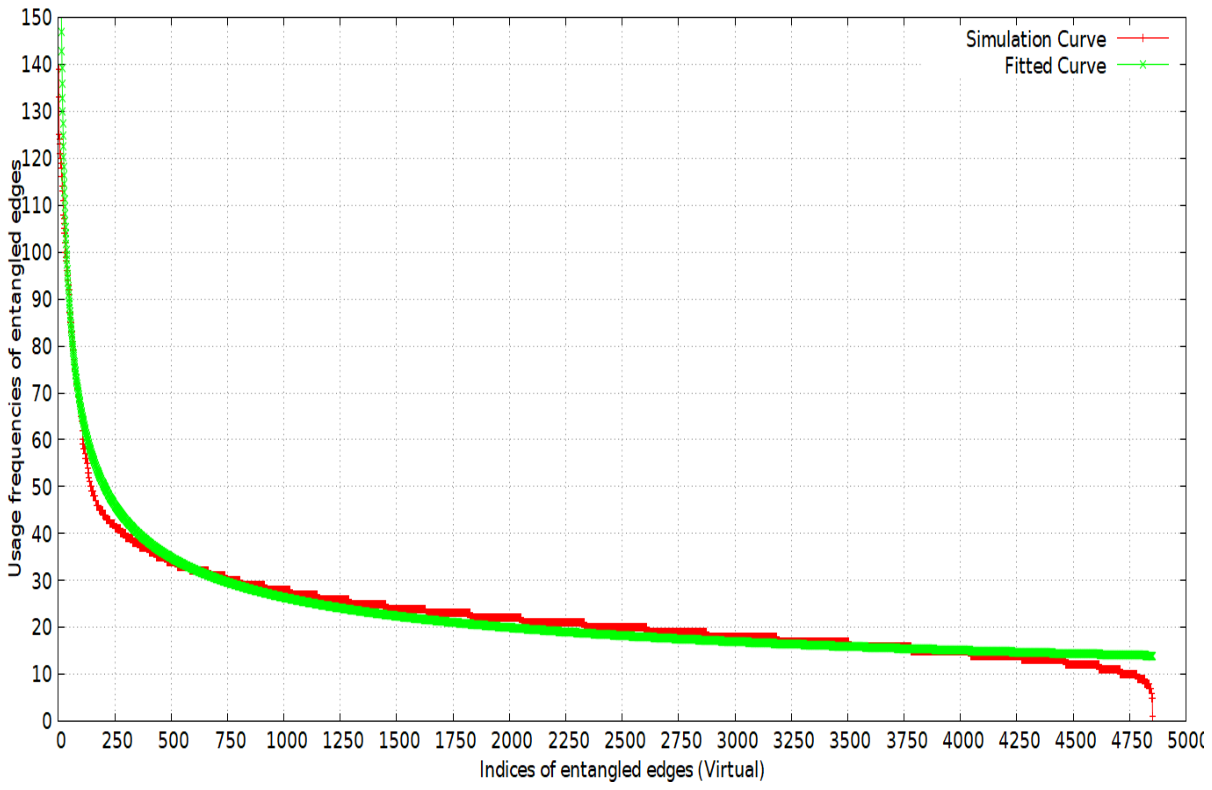}
\caption{Connection Setup - Power Law Distribution, virtual entanglements only, curve fit $f_{e^{(v)}_{i}} = 424.8808i^{-0.4023611}$ where $i$ is the virtual edge}
\label{plot_powerlaw_virtual_ent_edges_freq_vs_edge_indices_sorted_curve_fit_ink}
\end{figure}

\begin{figure}[ht]
\centering
\includegraphics[width=\columnwidth]{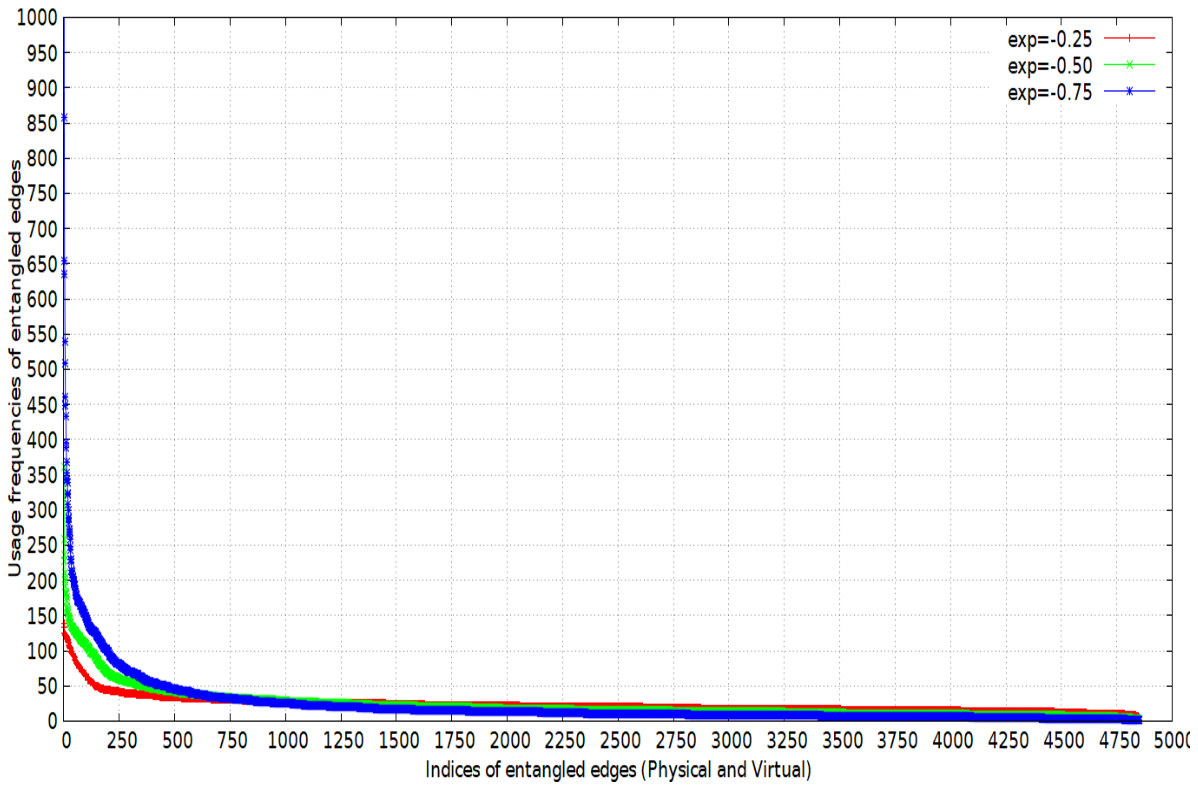}
\caption{Entangled edges (Physical and Virtual) usage with decreasing exponent}
\label{plot_script_powerlaw_decreasing_exp_all_ent_edges_freq_vs_edge_indices_sorted_ink}
\end{figure}

\begin{figure}[ht]
\centering
\includegraphics[width=\columnwidth]{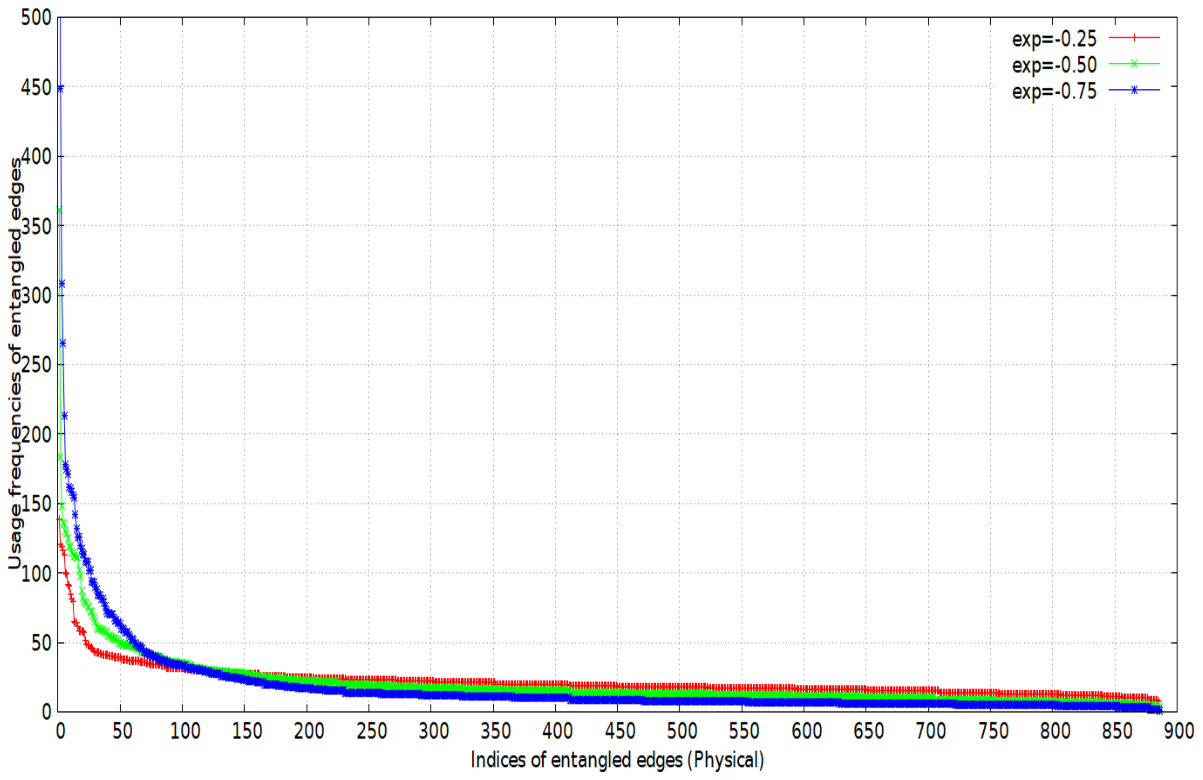}
\caption{Entangled edges (Physical) usage with decreasing exponent}
\label{plot_script_powerlaw_decreasing_exp_phy_ent_edges_freq_vs_edge_indices_sorted_ink}
\end{figure}

\begin{figure}[ht]
\centering
\includegraphics[width=\columnwidth]{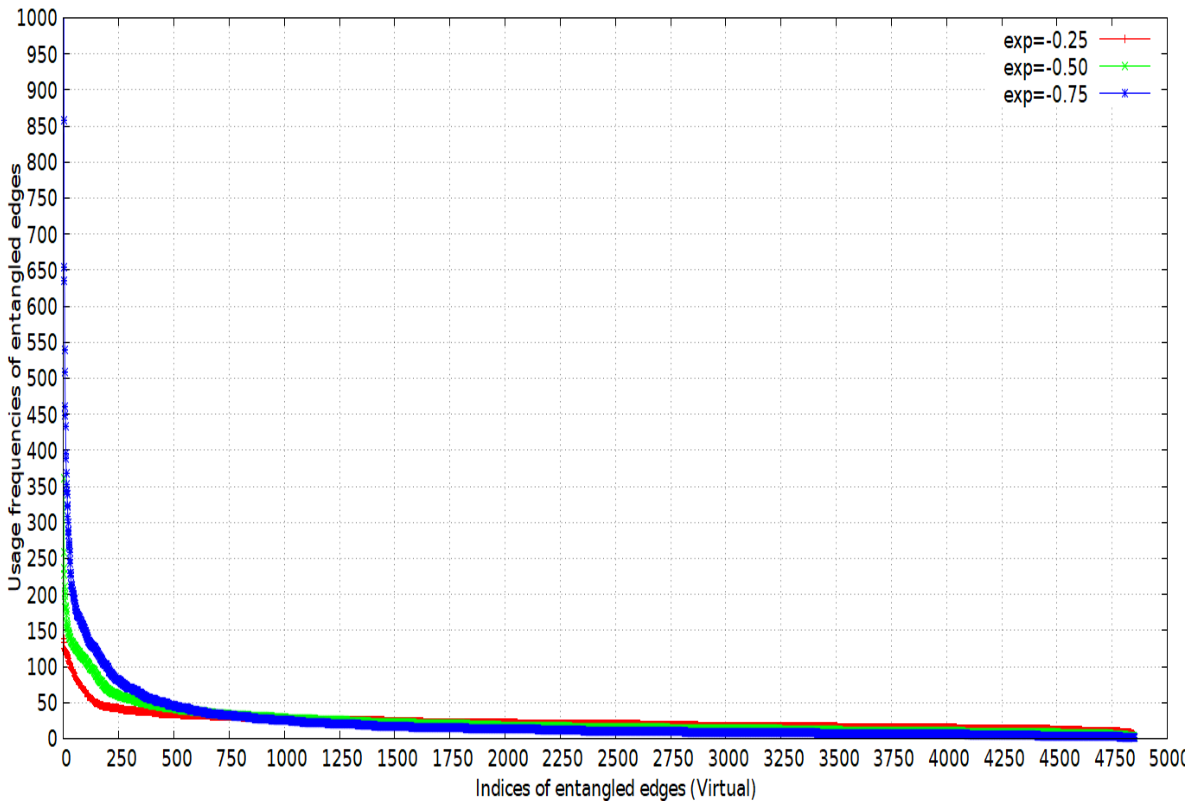}
\caption{Entangled edges (virtual) usage with decreasing exponent}
\label{plot_script_powerlaw_decreasing_exp_virtual_ent_edges_freq_vs_edge_indices_sorted_ink}
\end{figure}

If power law distribution is considered for connection setup, then as can be expected the entangled edges have frequencies which are also power law distributed with a coefficient of $424.8808$ and exponent of $-0.4023611$ as shown in Fig. \ref{plot_powerlaw_all_ent_edges_freq_vs_edge_indices_sorted_curve_fit_ink}. The physical edges (Fig. \ref{plot_powerlaw_phy_ent_edges_freq_vs_edge_indices_sorted_curve_fit_ink}) and virtual edges (Fig. \ref{plot_powerlaw_virtual_ent_edges_freq_vs_edge_indices_sorted_curve_fit_ink}) also show similar behaviours.
% decreasing exponent
Fig. \ref{plot_script_powerlaw_decreasing_exp_all_ent_edges_freq_vs_edge_indices_sorted_ink} shows the behaviour of edge usages for power law distribution of connection setups with decreasing exponents of $-0.25$, $-0.50$ and $-0.75$. It is observed that with lower exponents the centrality of edges increases. Similar behaviours are observed for both physical and virtual edges in Fig. \ref{plot_script_powerlaw_decreasing_exp_phy_ent_edges_freq_vs_edge_indices_sorted_ink} and Fig. \ref{plot_script_powerlaw_decreasing_exp_virtual_ent_edges_freq_vs_edge_indices_sorted_ink} respectively.
\subsubsection{Compare Uniform, Gaussian and Power Law}
\begin{figure}[ht]
\centering
\includegraphics[width=\columnwidth]{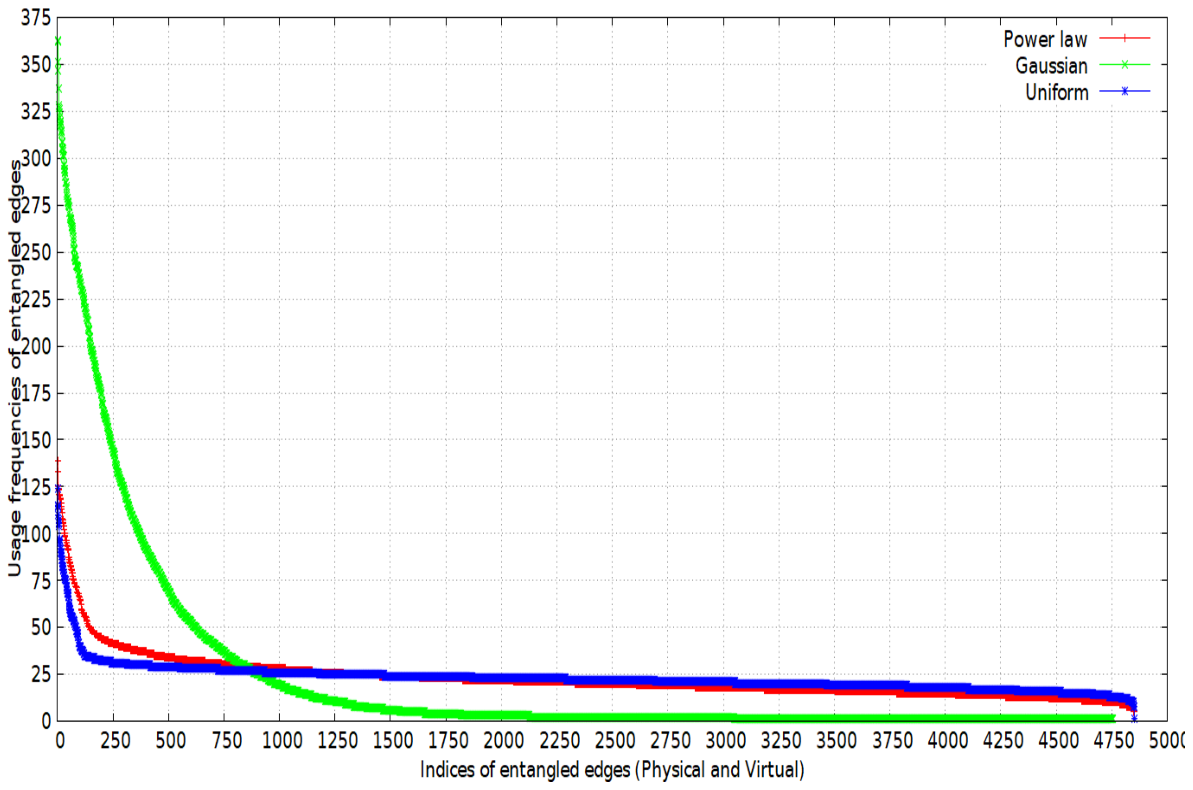}
\caption{Connection Setup - Compare Uniform, Normal and Power Law Distributions (Physical and Virtual)}
\label{plot_compare_uniform_norm_powerlaw_all_ent_edges_freq_vs_edge_indices_sorted_ink}
\end{figure}

\begin{figure}[ht]
\centering
\includegraphics[width=\columnwidth]{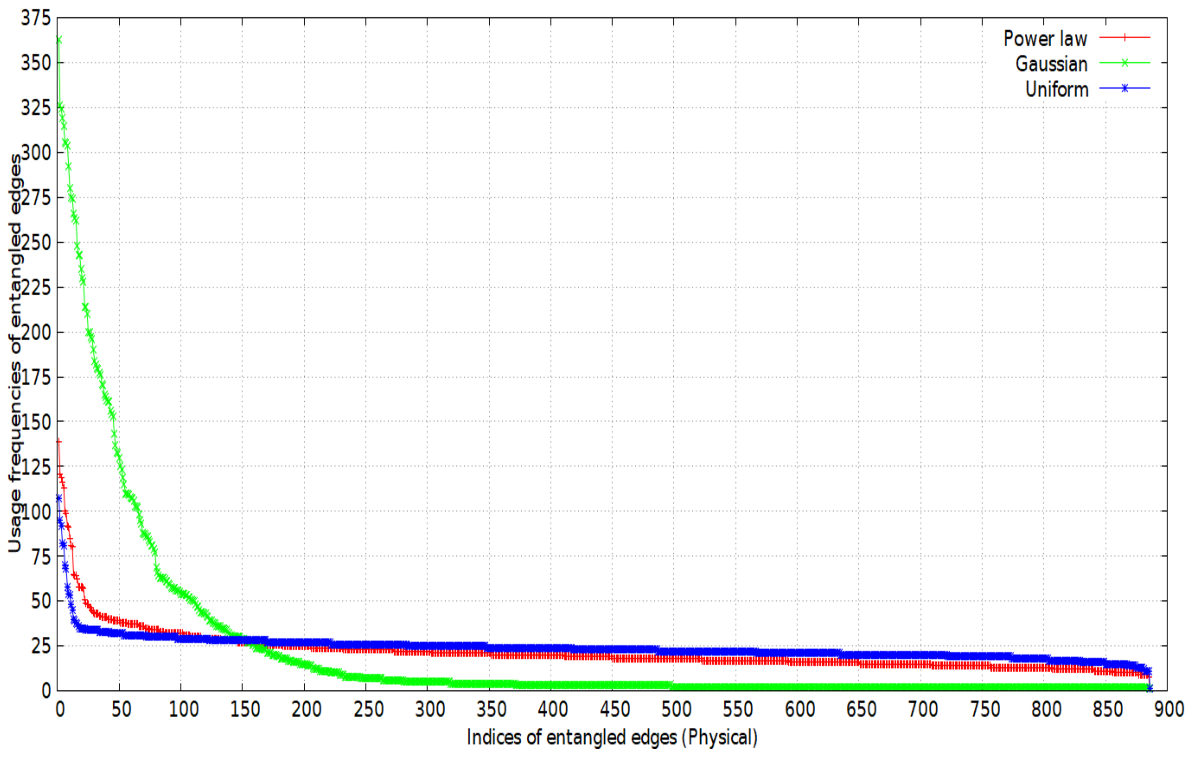}
\caption{Connection Setup - Compare Uniform, Normal and Power Law Distributions (Physical)}
\label{plot_compare_uniform_norm_powerlaw_phy_ent_edges_freq_vs_edge_indices_sorted_ink}
\end{figure}

\begin{figure}[ht]
\centering
\includegraphics[width=\columnwidth]{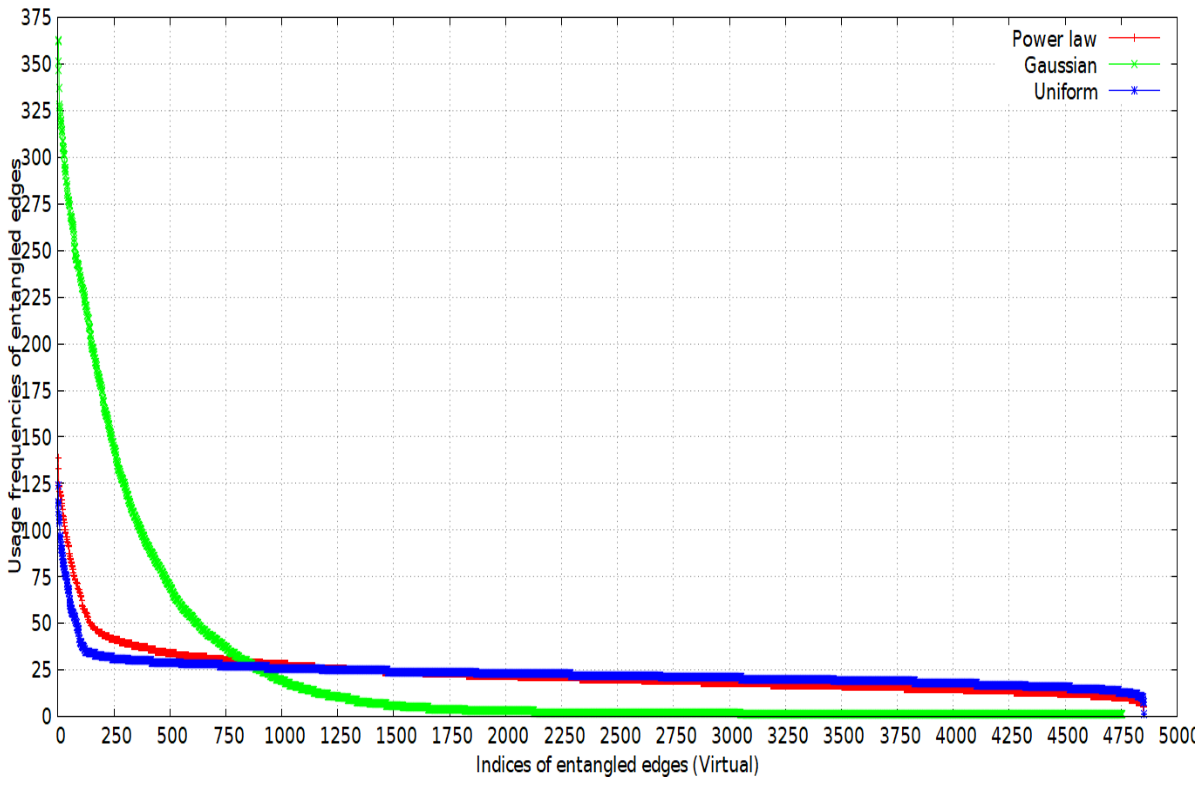}
\caption{Connection Setup - Compare Uniform, Normal and Power Law Distributions (Virtual)}
\label{plot_compare_uniform_norm_powerlaw_virtual_ent_edges_freq_vs_edge_indices_sorted_ink}
\end{figure}

Fig. \ref{plot_compare_uniform_norm_powerlaw_all_ent_edges_freq_vs_edge_indices_sorted_ink} compares the behaviour of usage of entangled edges for uniform, gaussian and power law distributions. For uniform and power law distributions, about 100 edges show higher centralities and the rest have lower usages. However, the gaussian distribution shows more edges having higher usages and the rest of them are close to zero akin to gaussian behaviour. Fig. \ref{plot_compare_uniform_norm_powerlaw_phy_ent_edges_freq_vs_edge_indices_sorted_ink} and Fig. \ref{plot_compare_uniform_norm_powerlaw_virtual_ent_edges_freq_vs_edge_indices_sorted_ink} respectively show similar behaviour for physical and virtual entangled edges.
\subsubsection{Entanglement build up with connections}
\begin{figure}[ht]
\centering
\includegraphics[width=\columnwidth]{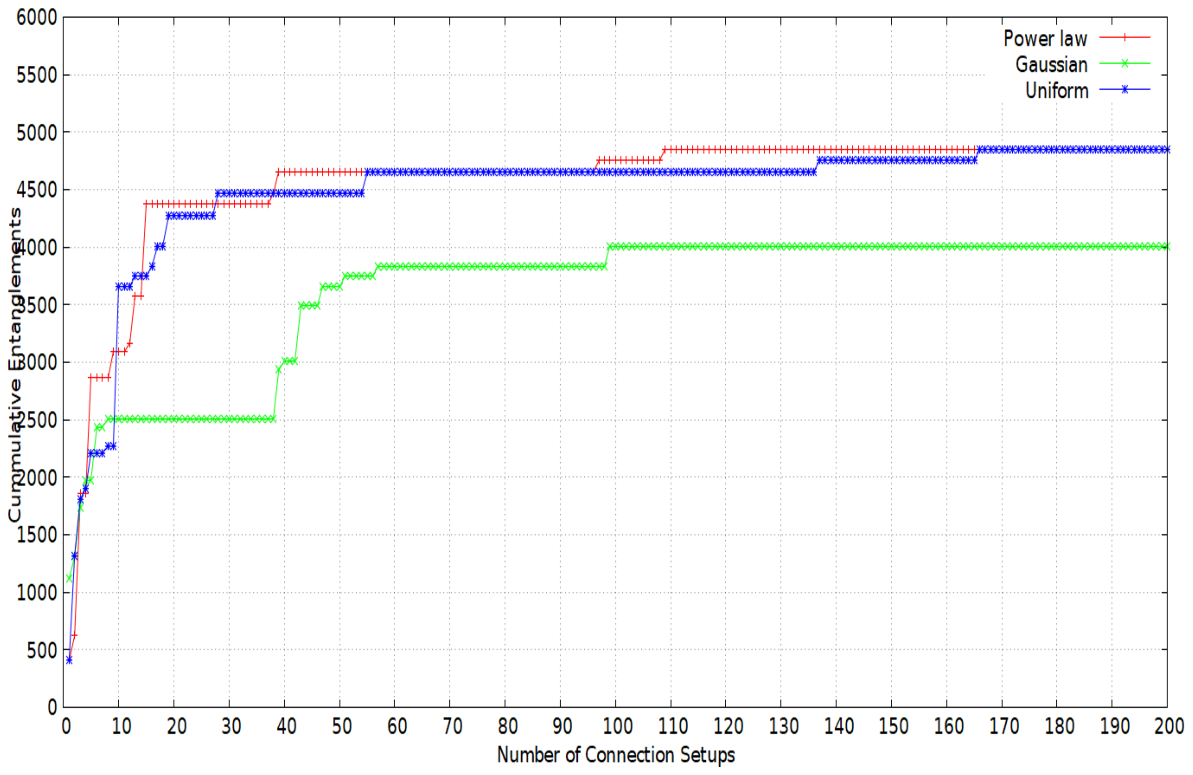}
\caption{Entanglement build up with connections - Compare Uniform, Normal and Power Law Distributions}
\label{plot_compare_uniform_norm_powerlaw_data_raw_qent_growth_vs_connection_ink}
\end{figure}

\begin{figure}[ht]
\centering
\includegraphics[width=\columnwidth]{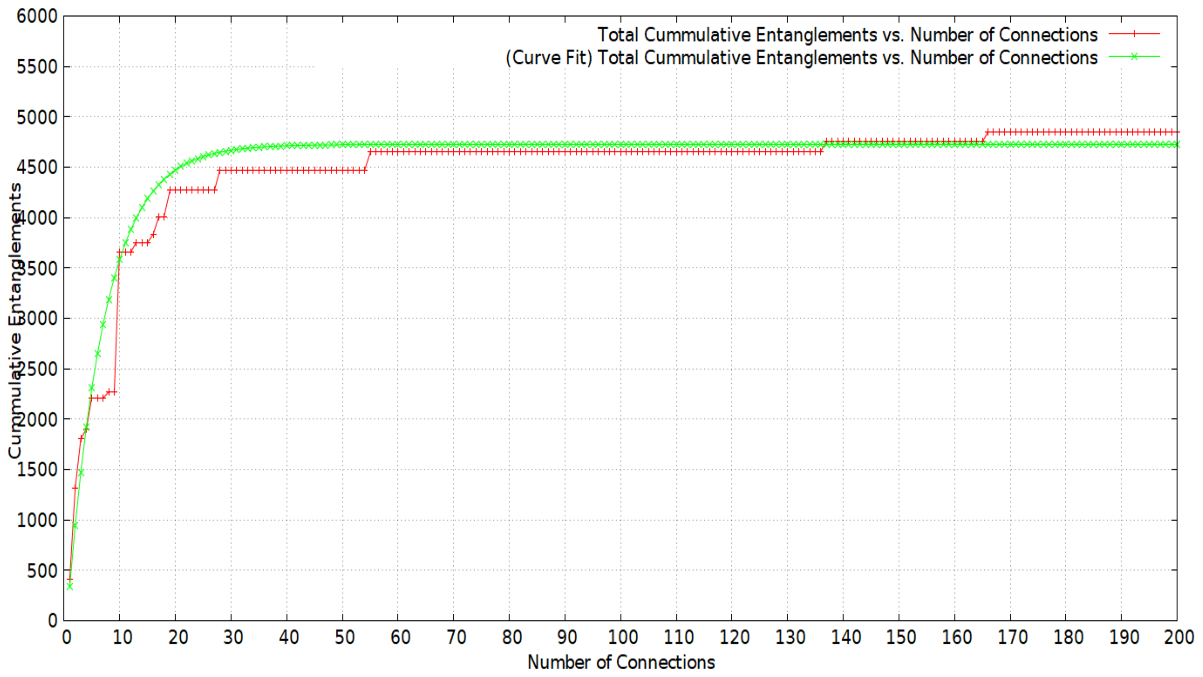}
\caption{Entanglement build up with connections - Uniform distribution. Curve fit $E^{(total)} = 4725-5100e^{-0.15k}$}
\label{plot_uniform_data_raw_qent_growth_vs_connection_ink}
\end{figure}

\begin{figure}[ht]
\centering
\includegraphics[width=\columnwidth]{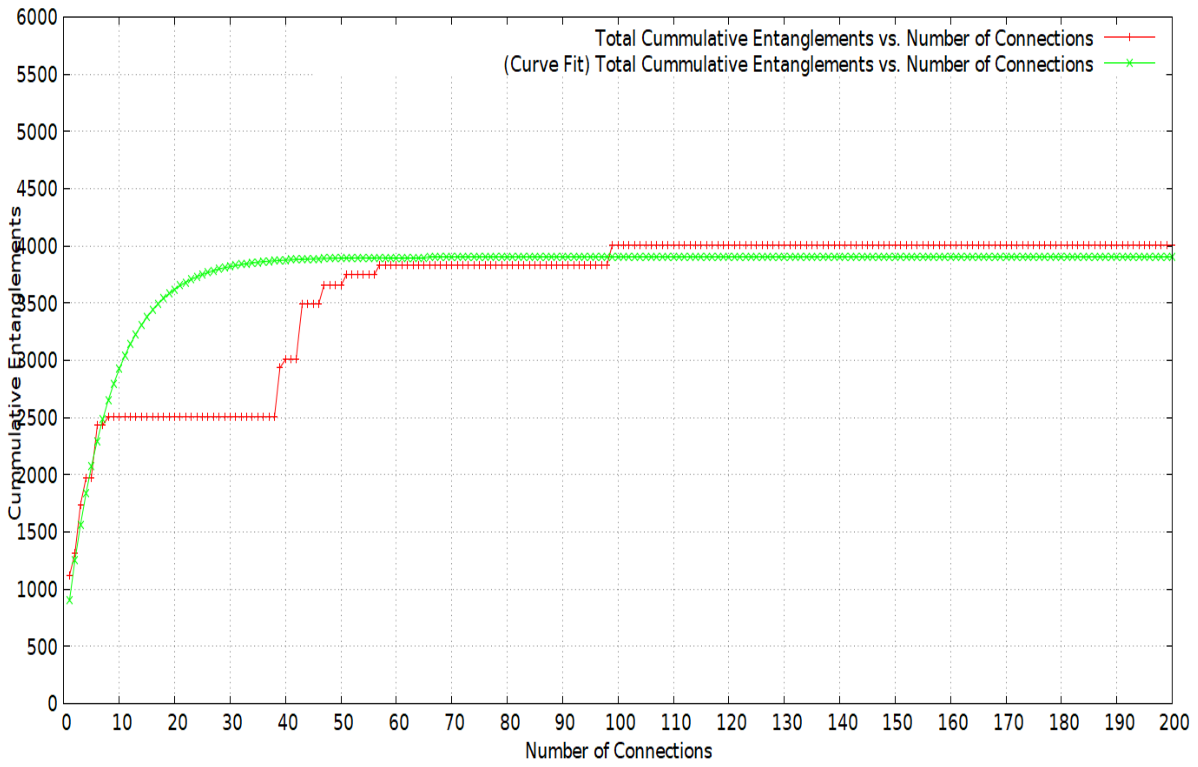}
\caption{Entanglement build up with connections with Gaussian sampling of end nodes. Curve fit $E^{(total)} = 3900-3400e^{-0.125k}$}
\label{plot_normal_data_raw_qent_growth_vs_connection_ink}
\end{figure}

\begin{figure}[ht]
\centering
\includegraphics[width=\columnwidth]{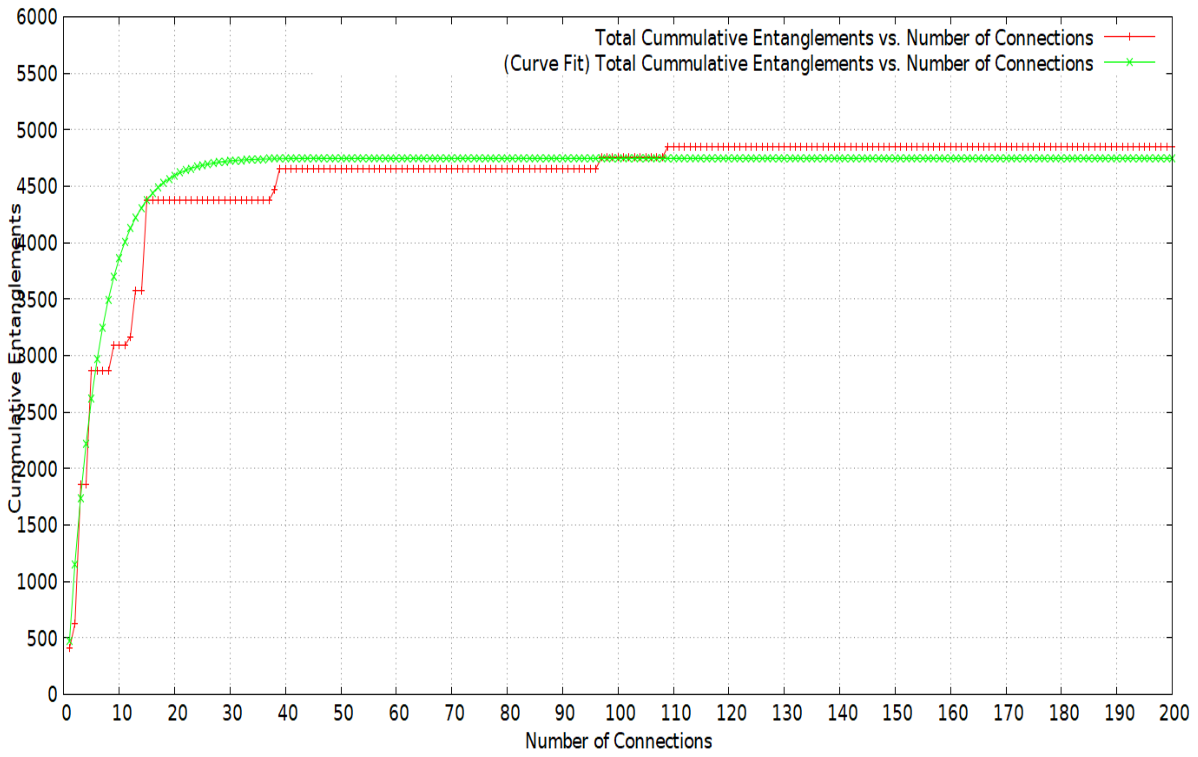}
\caption{Entanglement build up with connections with Power law sampling of end nodes. Curve fit $E^{(total)} = 4750-4750e^{-0.175k}$}
\label{plot_power_law_data_raw_qent_growth_vs_connection_ink}
\end{figure}

The number of entanglements that occur in the quantum network as connection setups are made following uniform, gaussian and power law distributions for end node selections is shown in Fig. \ref{plot_compare_uniform_norm_powerlaw_data_raw_qent_growth_vs_connection_ink}. While \emph{x}-axis shows the number of connections, \emph{y}-axis is the cummulative number of entanglements. Though, there are 100,000 connections in the simulation, it is observed that the entanglements build up to their maximum value within the first 200 connections. Hence, all the connections are not shown. It is observed that the maximum number of entanglements is almost the same for power law and uniform distributions but lower for gaussian. Also, power law reaches its highest value first, quickly followed by the gaussian and much later by the uniform distribution. All three growths can be approximated with monomolecular function. Figs. \ref{plot_uniform_data_raw_qent_growth_vs_connection_ink}, \ref{plot_normal_data_raw_qent_growth_vs_connection_ink} and \ref{plot_power_law_data_raw_qent_growth_vs_connection_ink} show the growths for uniform, normal and power law distribution connection setups along with their fitted monomolecular curve fits. For power law and uniform, the curve fits are better whereas the gaussian has a gap in between but matches well in most parts of the graph. A composite function of multiple monomolecular curves may fit better for the gaussian case though it will be complex.

\begin{figure}[ht]
\centering
\includegraphics[width=\columnwidth]{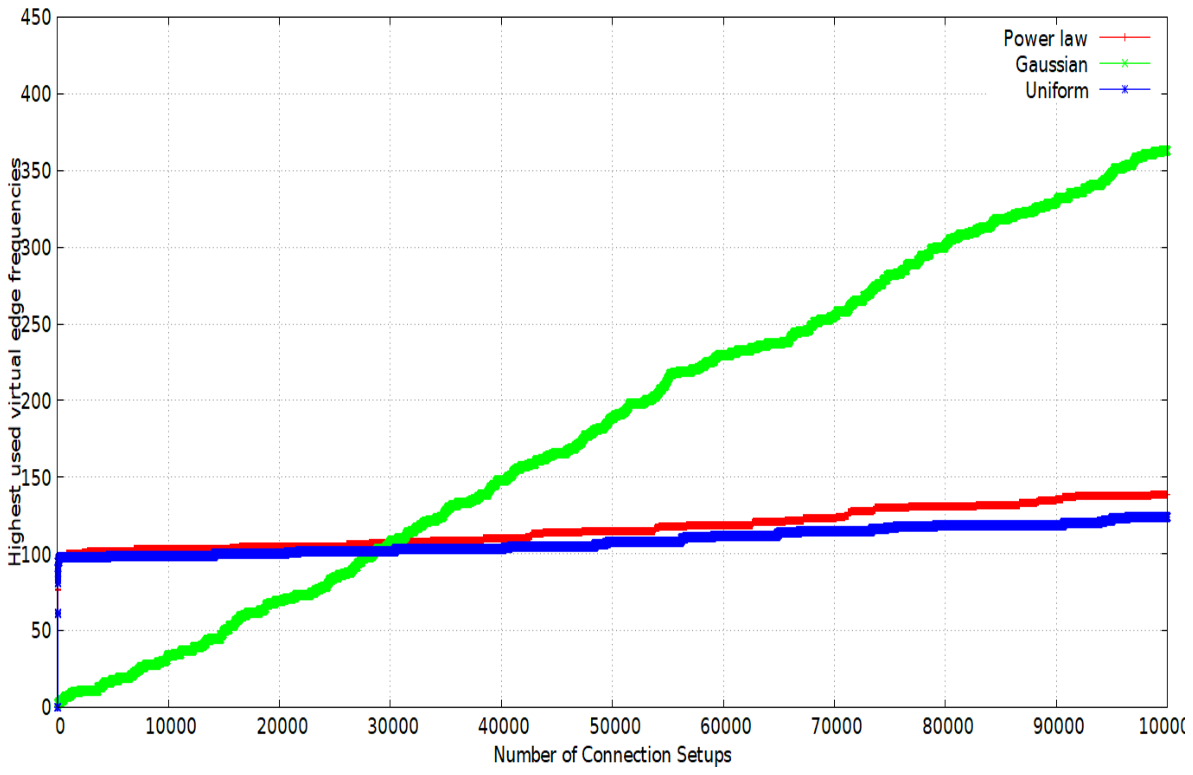}
\caption{Frequencies of highest used virtual edge - Compare Uniform, Normal and Power Law Distributions}
\label{plot_compare_uniform_norm_powerlaw_data_highest_freq_edge_vs_connection_ink}
\end{figure}

\begin{figure}[ht]
\centering
\includegraphics[width=\columnwidth]{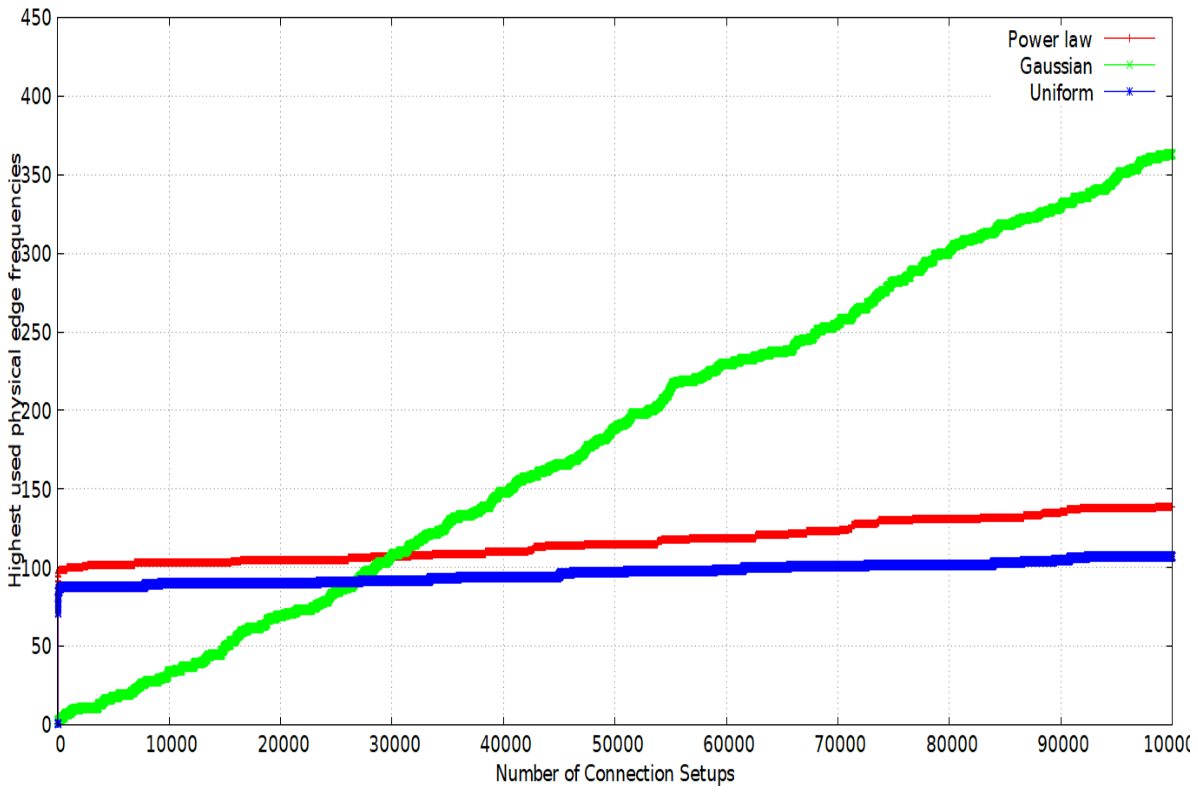}
\caption{Frequencies of highest used physical edge - Compare Uniform, Normal and Power Law Distributions}
\label{plot_compare_uniform_norm_powerlaw_data_highest_freq_phy_edge_vs_connection_ink}
\end{figure}

Fig. \ref{plot_compare_uniform_norm_powerlaw_data_highest_freq_edge_vs_connection_ink} shows the usage frequency of the highest used virtual edge, i.e., $max(f_{e^{(v)}_i})$ for all the three distributions. For gaussian, the growth is almost linear throughout. However, for the uniform and power law cases, there is an initial spike and then there is a gradual linear increase. Similar behaviour is also observed for the highest used physical edge, i.e, $max(f_{e^{(p)}_i})$, in Fig. \ref{plot_compare_uniform_norm_powerlaw_data_highest_freq_phy_edge_vs_connection_ink} for all the three cases. The only difference is the growth for power law is more than that of uniform for the physical edge.
\subsection{Node properties}
\begin{figure}[ht]
\centering
\includegraphics[width=\columnwidth]{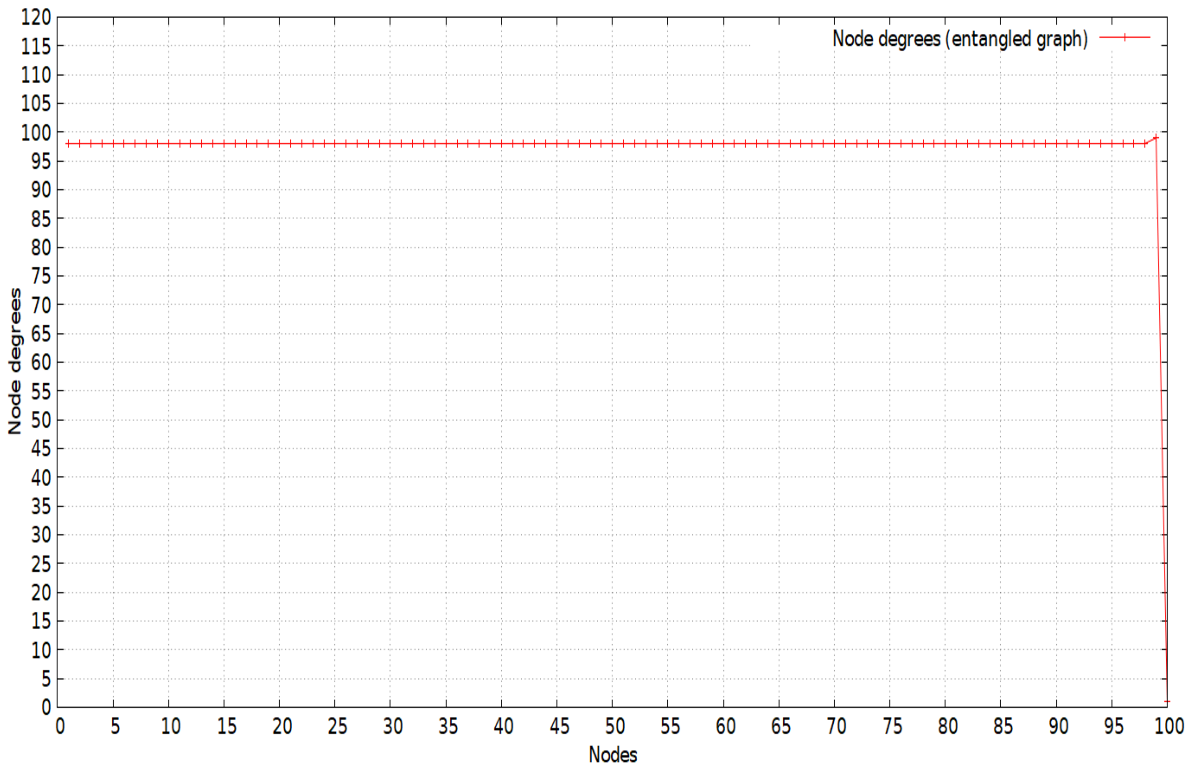}
\caption{Uniform - Degree of nodes with entanglements}
\label{plot_uniform_data_node_degree_ent_graph_ink}
\end{figure}

\begin{figure}[ht]
\centering
\includegraphics[width=\columnwidth]{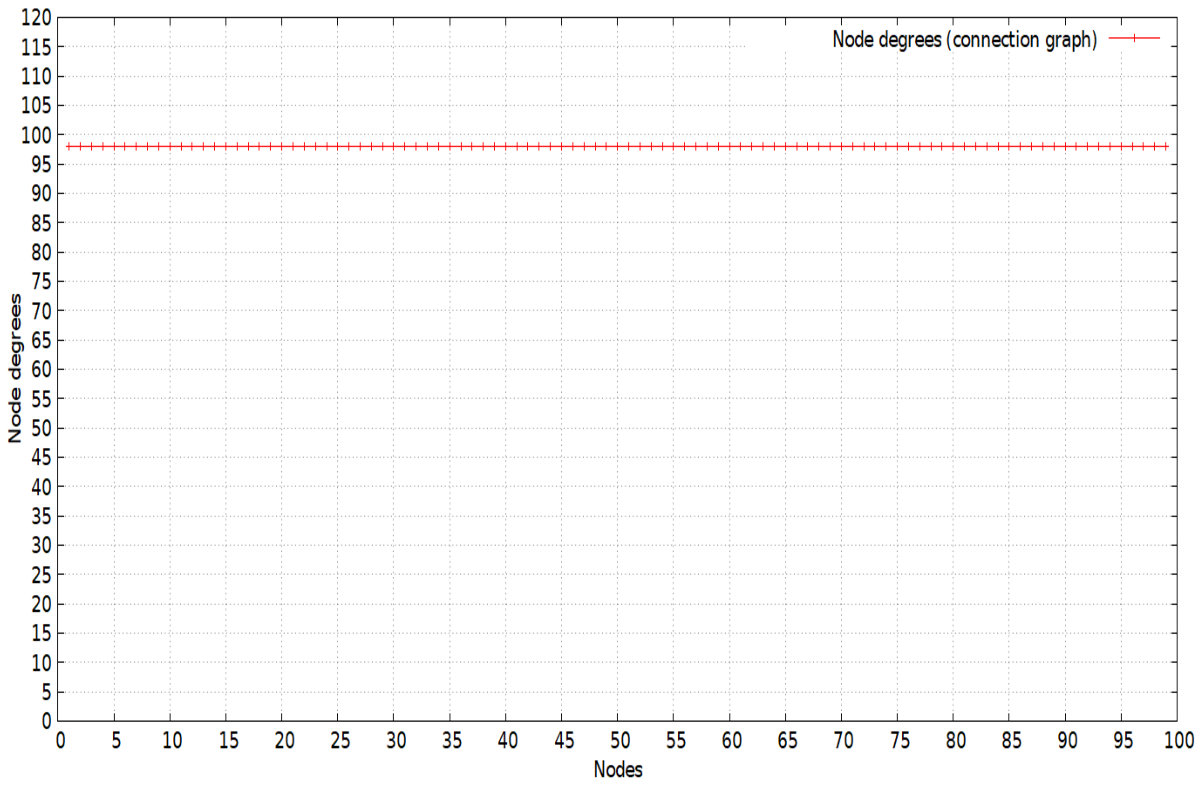}
\caption{Uniform - Degree of nodes with connections}
\label{plot_uniform_data_node_degree_connection_request_graph_ink}
\end{figure}

\begin{figure}[ht]
\centering
\includegraphics[width=\columnwidth]{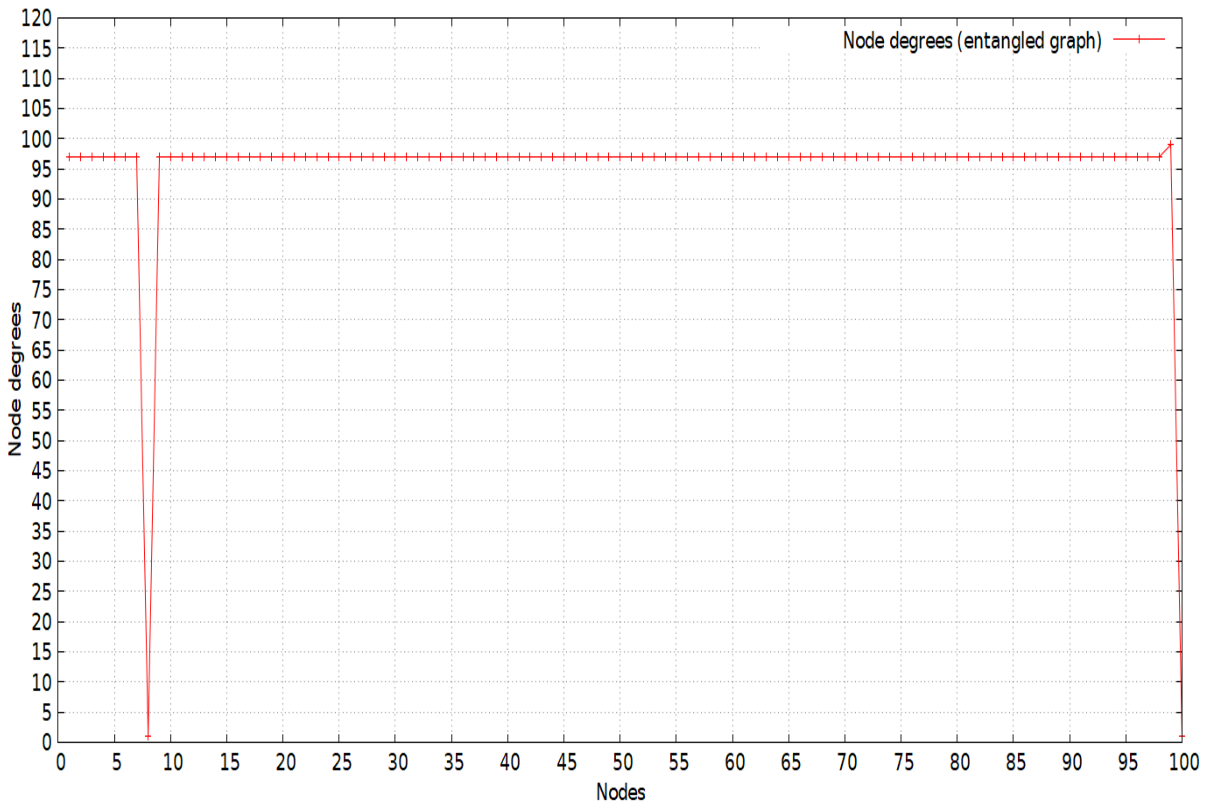}
\caption{Gaussian - Degree of nodes with entanglements}
\label{plot_normal_data_node_degree_ent_graph_ink}
\end{figure}

\begin{figure}[ht]
\centering
\includegraphics[width=\columnwidth]{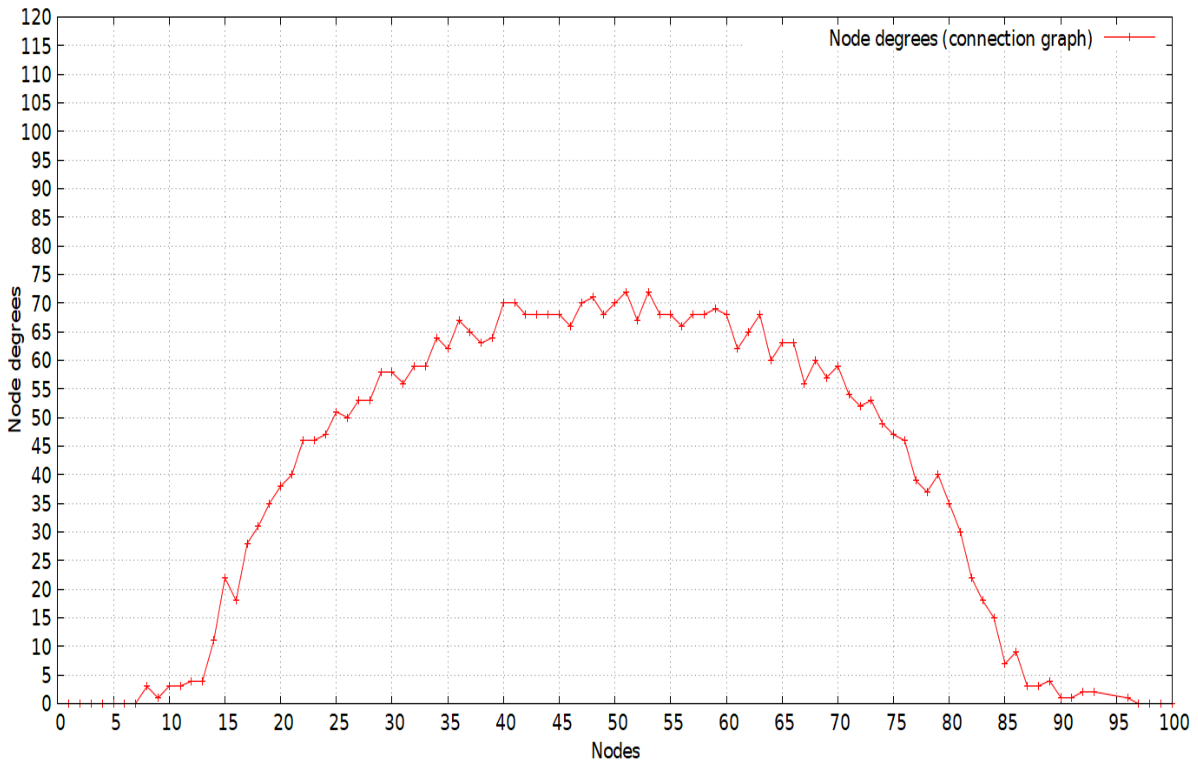}
\caption{Gaussian - Degree of nodes with connections}
\label{plot_normal_data_node_degree_connection_request_graph_ink}
\end{figure}

\begin{figure}[ht]
\centering
\includegraphics[width=\columnwidth]{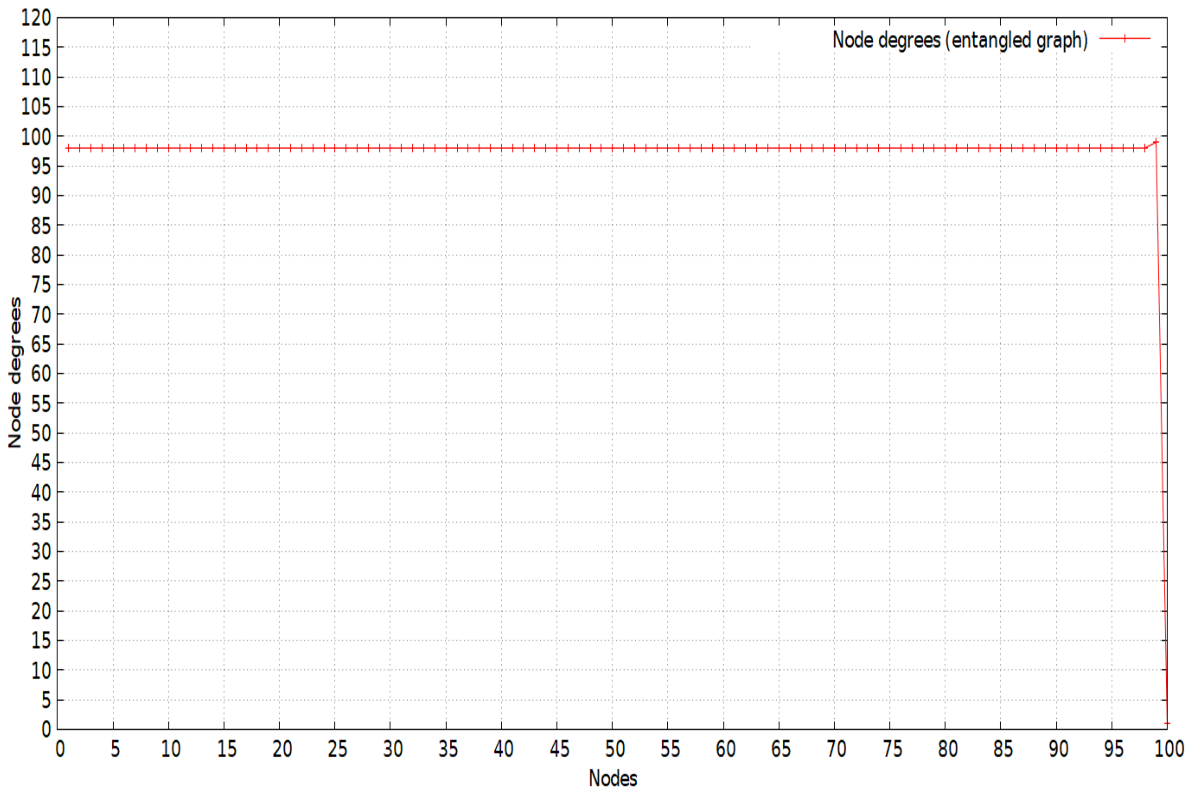}
\caption{Power law - Degree of nodes with entanglements}
\label{plot_power_law_data_node_degree_ent_graph_ink}
\end{figure}

\begin{figure}[ht]
\centering
\includegraphics[width=\columnwidth]{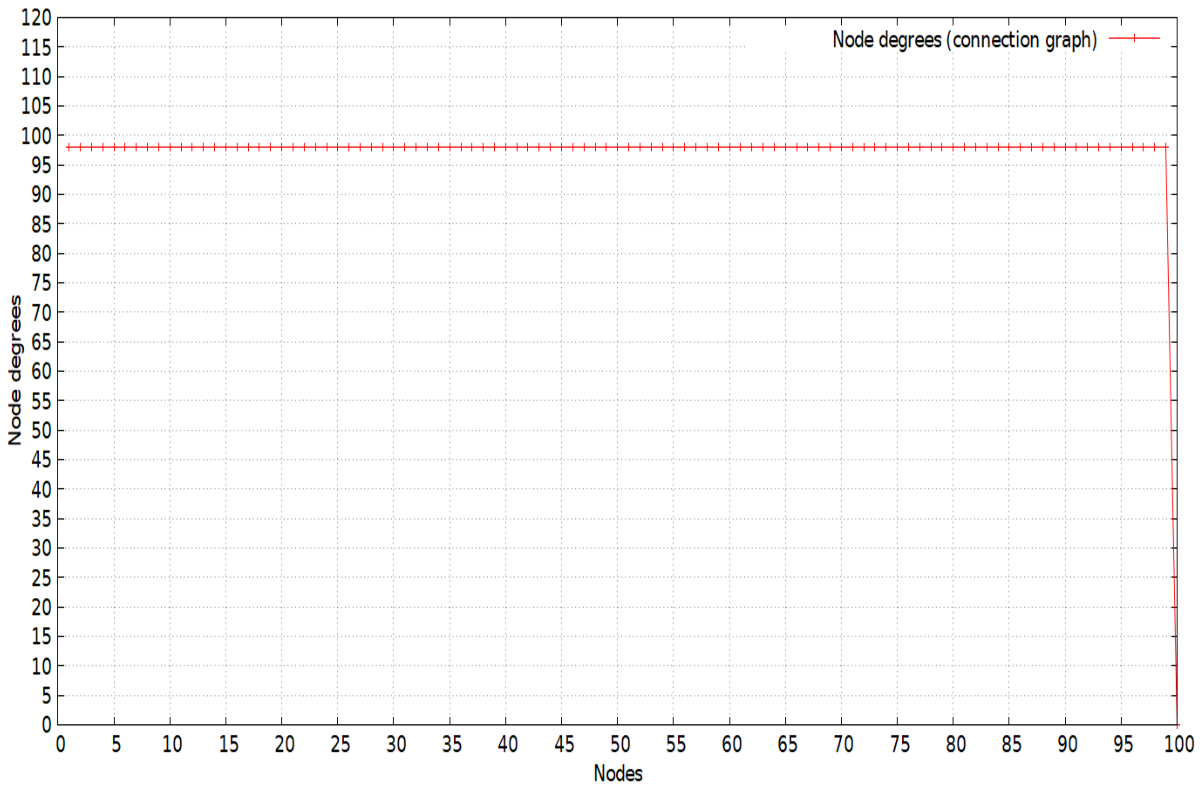}
\caption{Power law - Degree of nodes with connections}
\label{plot_power_law_data_node_degree_connection_request_graph_ink}
\end{figure}

This section presents the node centrality properties when the connection graph is uniform, gaussian and power law distributed. The final degrees distribution entangled graphs for the three scenarios are shown in Figs. \ref{plot_uniform_data_node_degree_ent_graph_ink}, \ref{plot_normal_data_node_degree_ent_graph_ink} and \ref{plot_power_law_data_node_degree_ent_graph_ink} respectively  after $M=100,000$ connections. The degree distribution of the corresponding connection graphs is shown in Figs. \ref{plot_uniform_data_node_degree_connection_request_graph_ink}, \ref{plot_normal_data_node_degree_connection_request_graph_ink} and \ref{plot_power_law_data_node_degree_connection_request_graph_ink} respectively. If the two end nodes of a connection are drawn from uniform and power law distributions, the degree of the nodes for both entangled and connection graphs are more or less uniform. This is because in both the cases all the nodes participate (at least once) during large number of connection setups which in turn leads to uniform entanglement distribution. However, the entangled graph for gaussian distributed connection setups has a bowl shaped form. This is due to the fact that some of the nodes do not participate in connection setups (owing to gaussian behaviour of near zero probability of selecting some nodes) and hence have zero degrees in the entangled graph. 

\begin{figure}[ht]
\centering
\includegraphics[width=\columnwidth]{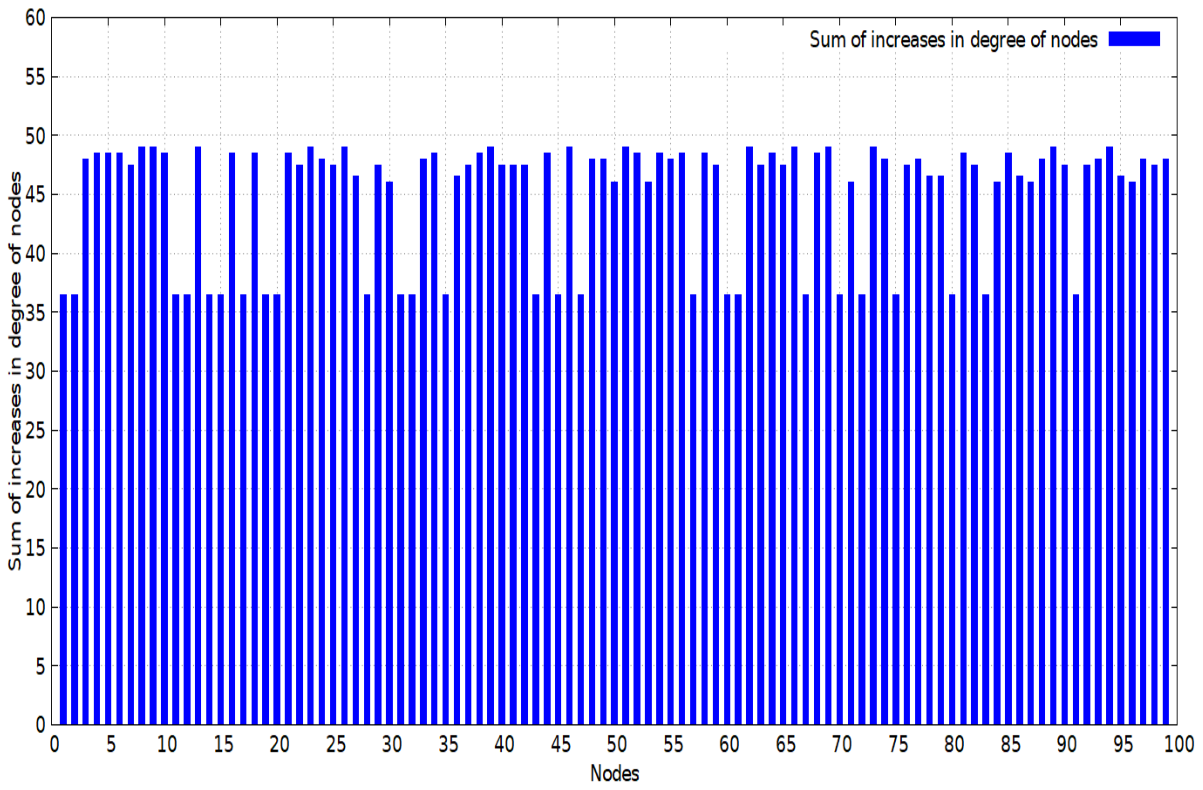}
\caption{Uniform - Increases in node degree growth with connections}
\label{plot_node_degree_growth_vs_connection_ink}
\end{figure}

\begin{figure}[ht]
\centering
\includegraphics[width=\columnwidth]{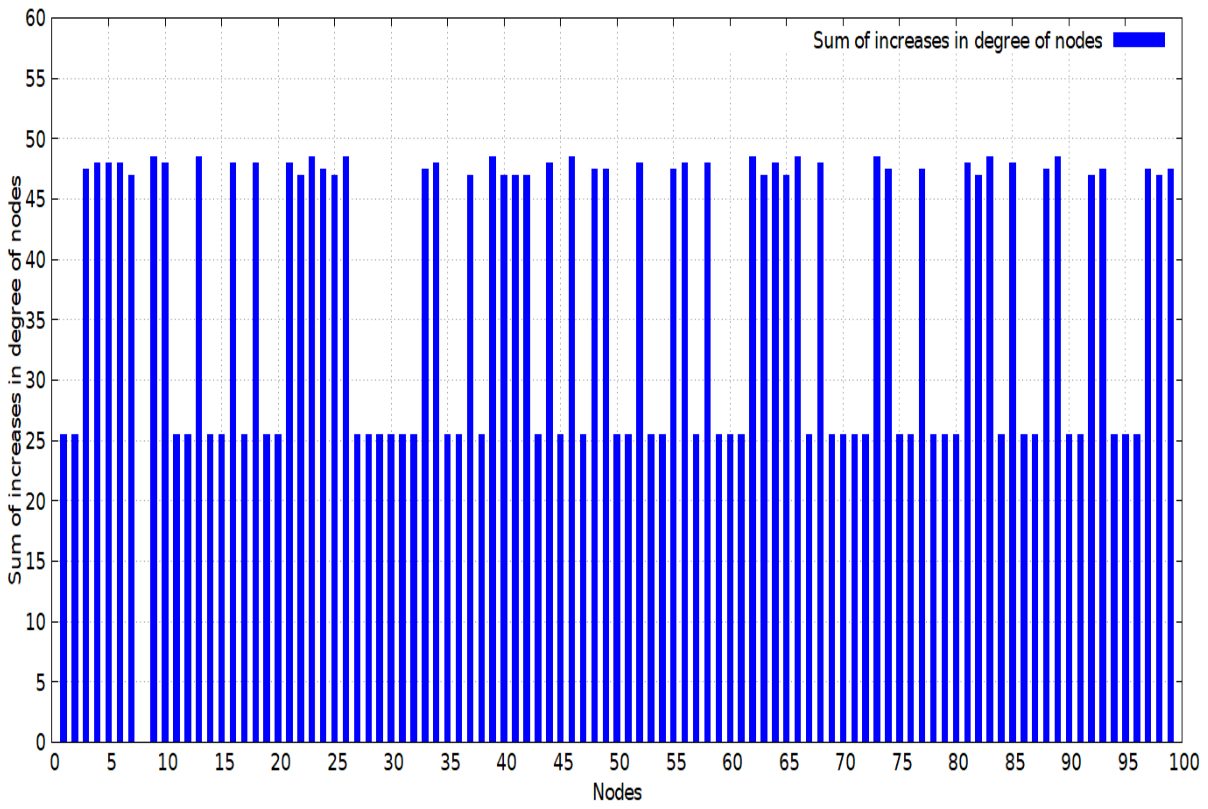}
\caption{Gaussian - Increases in node degree growth with connections}
\label{plot_normal_node_degree_growth_vs_connection_ink}
\end{figure}

\begin{figure}[ht]
\centering
\includegraphics[width=\columnwidth]{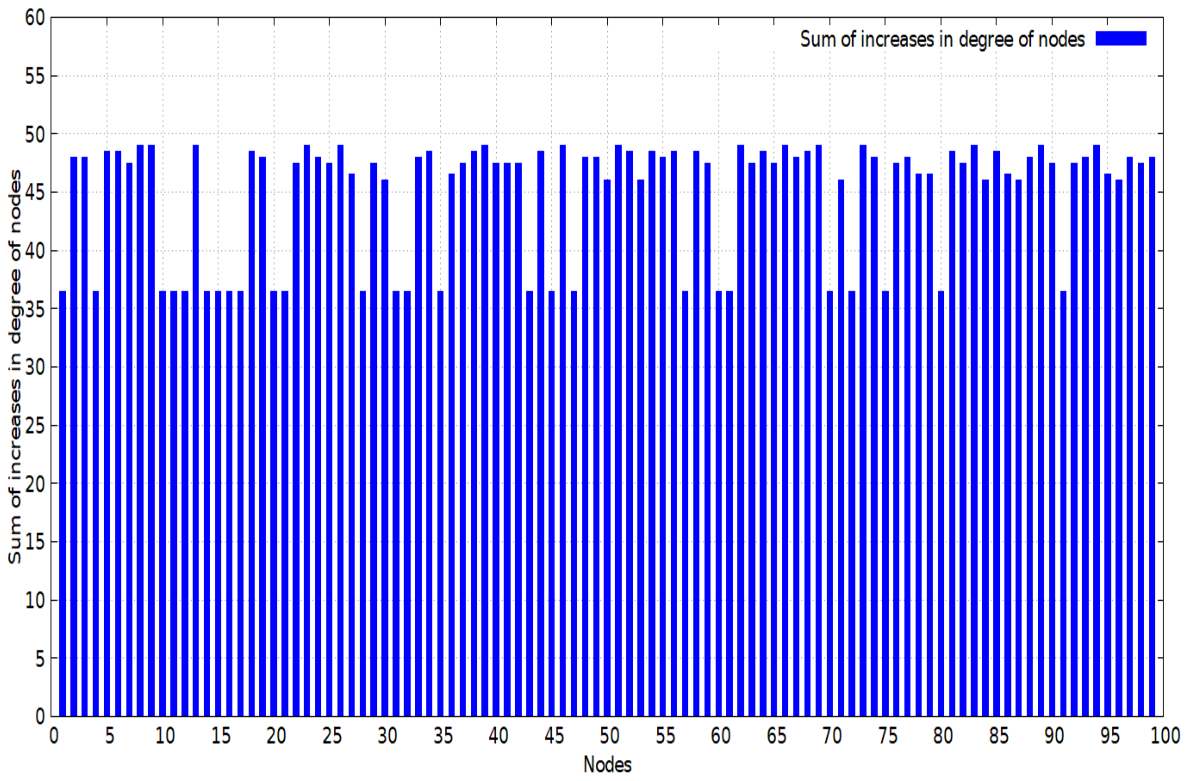}
\caption{Power law - Increases in node degree growth with connections}
\label{plot_power_law_node_degree_growth_vs_connection_ink}
\end{figure}

The growth of node degrees of the entangled graphs is studied next. The sum of growth in the degree of nodes, i.e.,  $\Delta d_j = \sum_{k=1}^{M} [d_j^{(k+1)} - d_j^{(k)}]$ for the three distributions of connection graphs is shown in Figs. \ref{plot_node_degree_growth_vs_connection_ink}, \ref{plot_normal_node_degree_growth_vs_connection_ink} and \ref{plot_power_law_node_degree_growth_vs_connection_ink} respectively.  Interestingly, two levels are clearly seen. Some of the nodes have low proactive entanglements in the beginning so they have higher changes and vice versa in the bar charts.

\begin{figure}[ht]
\centering
\includegraphics[width=\columnwidth]{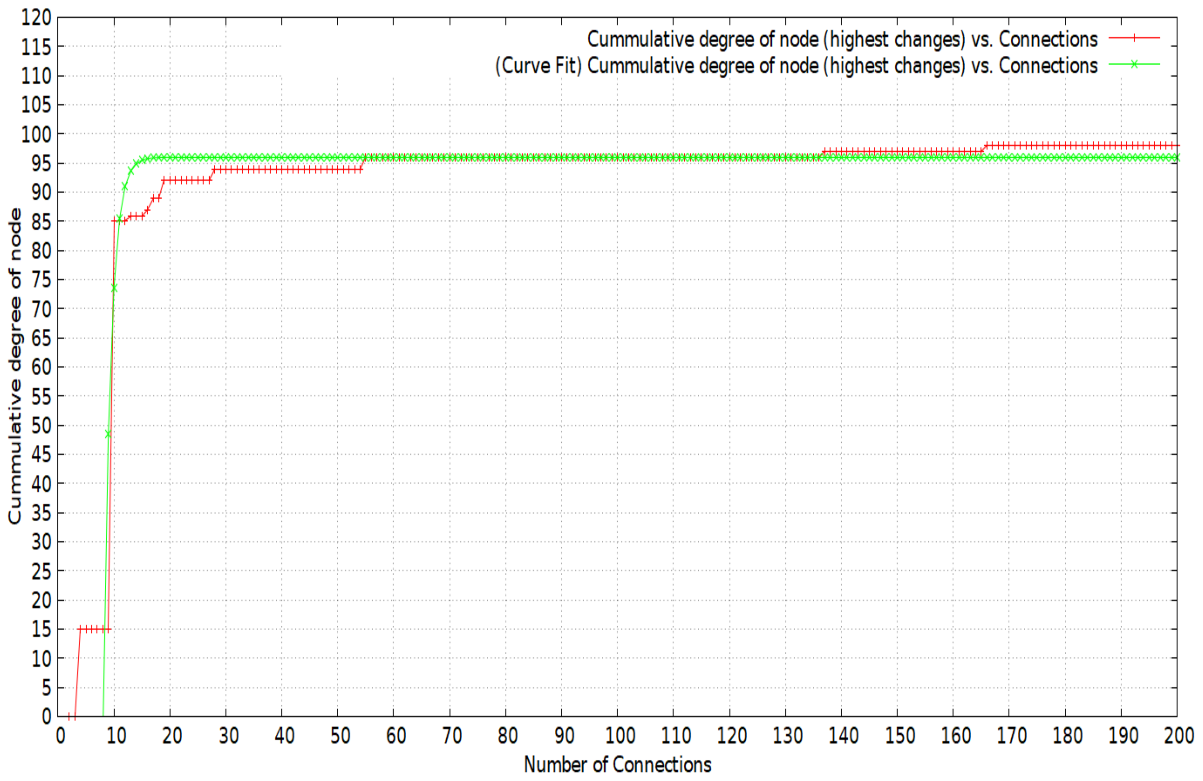}
\caption{Uniform - Cummulative degree of the node with highest changes vs. connection. Curve fit $d_j^{(k)} = 96-450e^{-0.75k}$ for $k>6$, $0$ otherwise }
\label{plot_node_highest_degree_growth_vs_connection_ink}
\end{figure}

\begin{figure}[ht]
\centering
\includegraphics[width=\columnwidth]{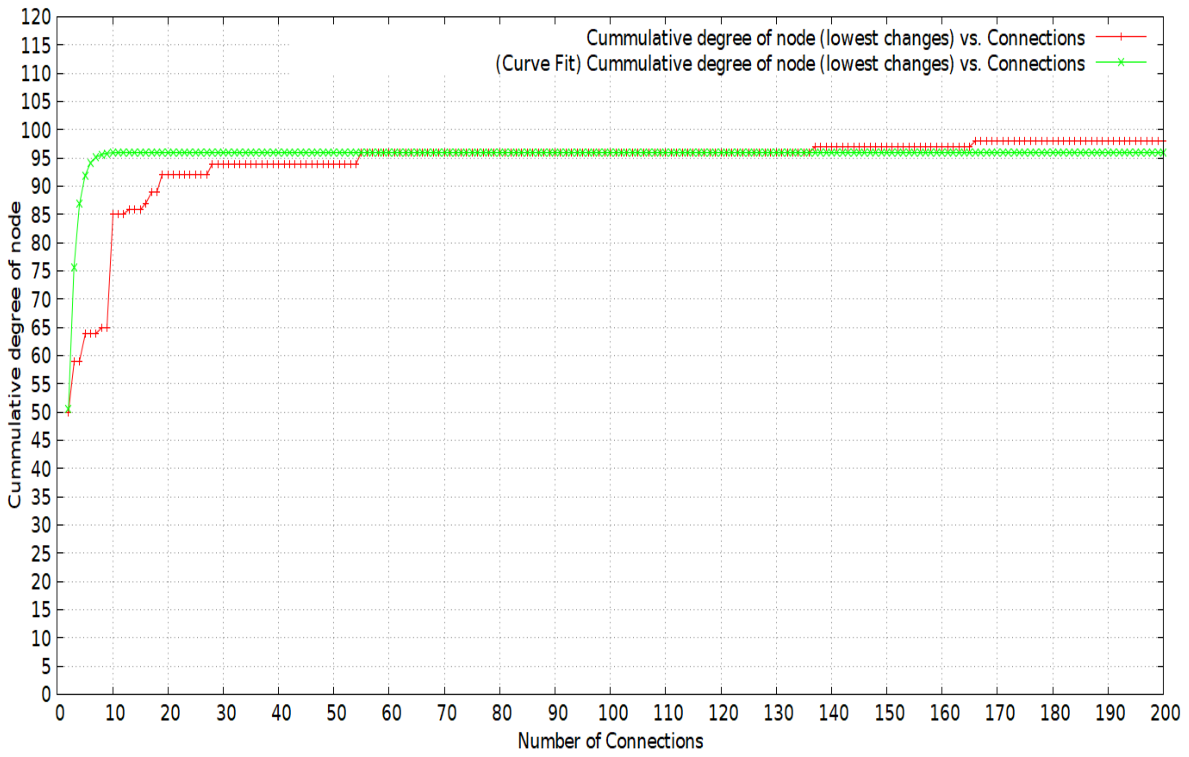}
\caption{Uniform - Cummulative degree of the node with highest changes vs. connection. Curve fit $d_j^{(k)} = 96-225e^{-0.8k}$}
\label{plot_node_lowest_degree_growth_vs_connection_ink}
\end{figure}

\begin{figure}[ht]
\centering
\includegraphics[width=\columnwidth]{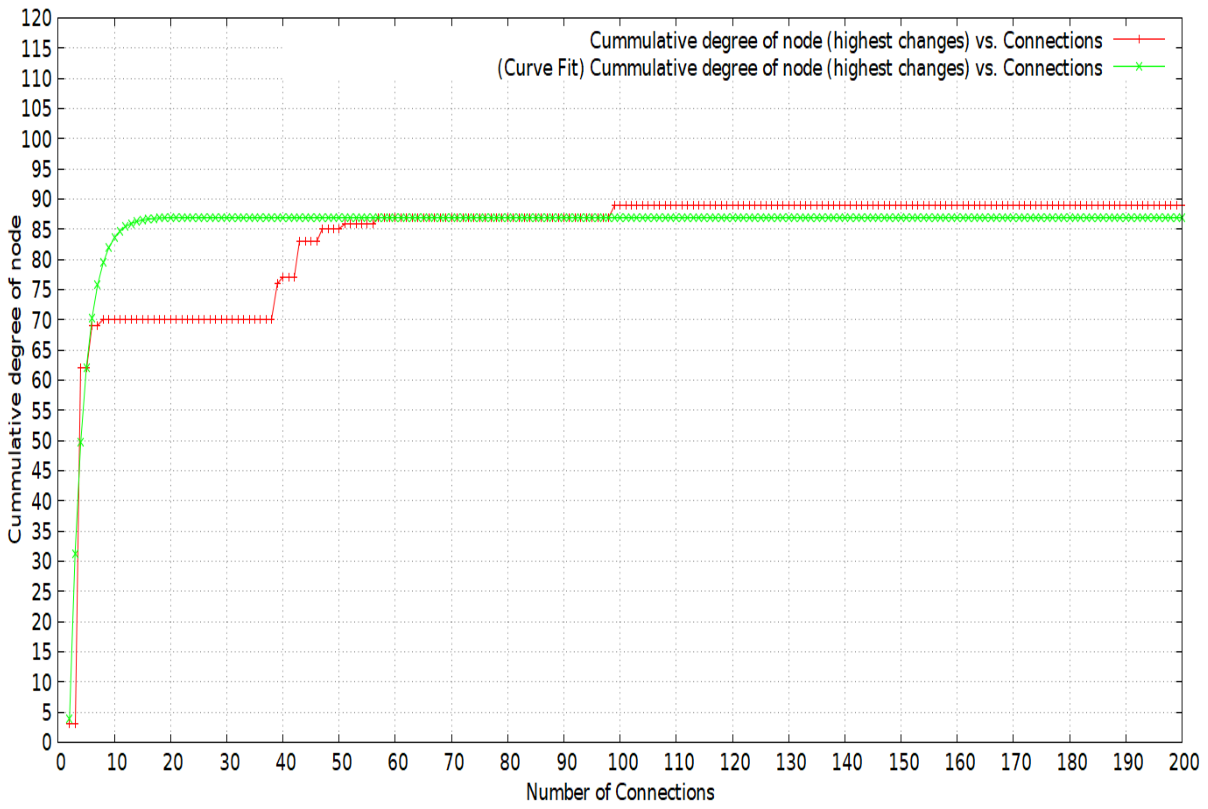}
\caption{Gaussian - Cummulative degree of the node with highest changes vs. connection. Curve fit $d_j^{(k)} = 87-185e^{-0.40k}$}
\label{plot_normal_node_highest_degree_growth_vs_connection_ink}
\end{figure}

\begin{figure}[ht]
\centering
\includegraphics[width=\columnwidth]{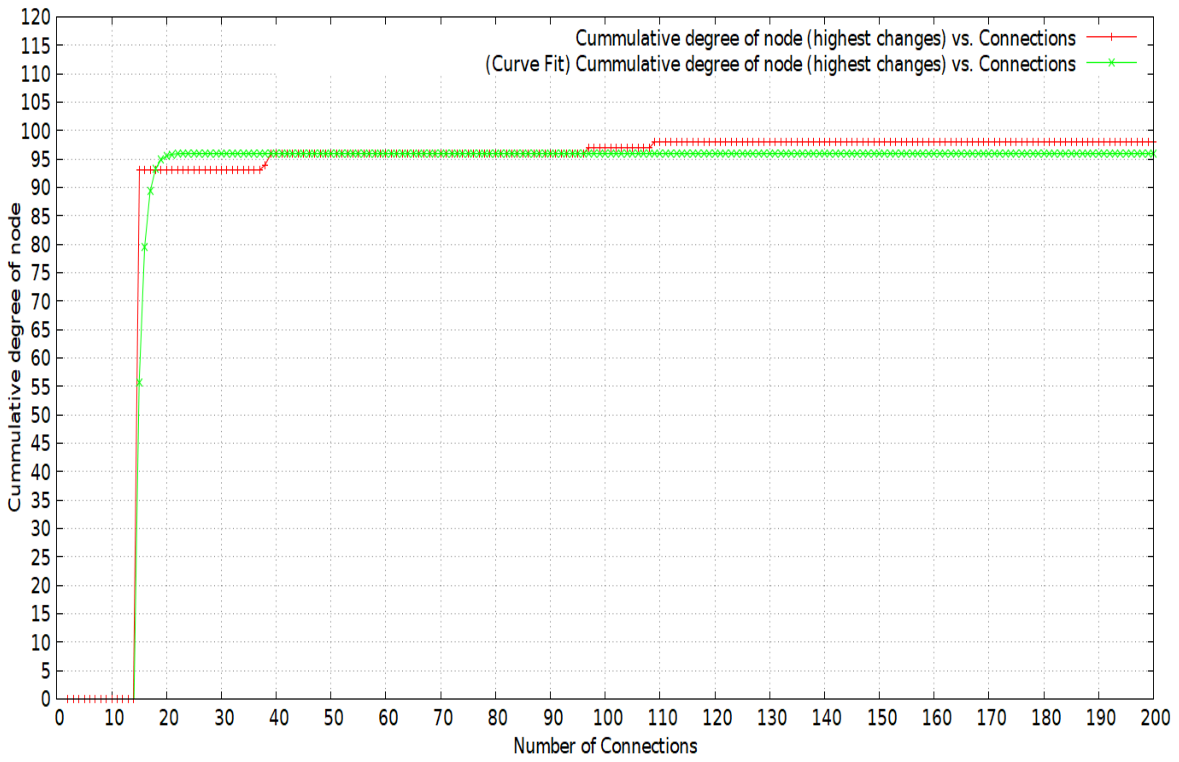}
\caption{Power law - Cummulative degree of the node with highest changes vs. connection. Curve fit $d_j^{(k)} = 96-600e^{-0.90k}$}
\label{plot_power_law_node_highest_degree_growth_vs_connection_ink}
\end{figure}

\begin{figure}[ht]
\centering
\includegraphics[width=\columnwidth]{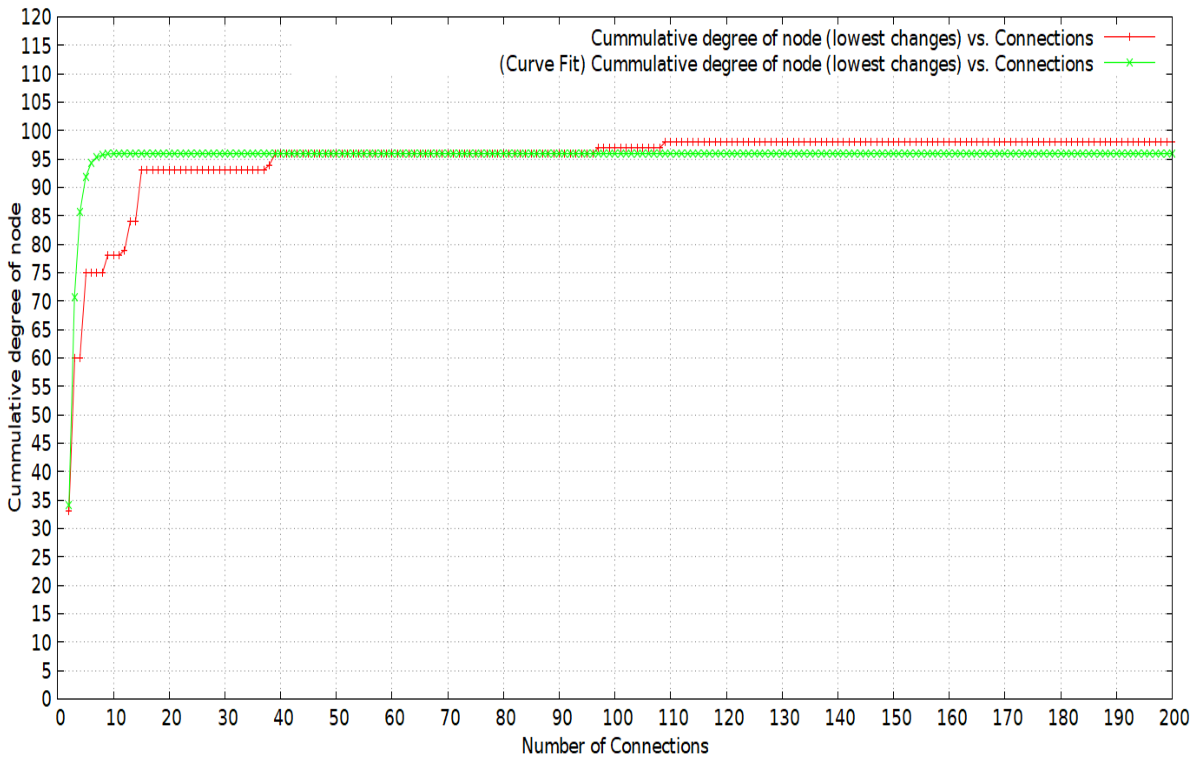}
\caption{Power law - Cummulative degree of the node with lowest changes vs. connection. Curve fit $d_j^{(k)} = 96-375e^{-0.90k}$}
\label{plot_power_law_node_lowest_degree_growth_vs_connection_ink}
\end{figure}

The node with the highest changes $max{(\Delta d_j)}$ for the three distributions along with the curve fits are shown in Fig. \ref{plot_node_highest_degree_growth_vs_connection_ink}, (\ref{plot_normal_node_highest_degree_growth_vs_connection_ink}) and \ref{plot_power_law_node_highest_degree_growth_vs_connection_ink}. Corresponding plots for $min{(\Delta d_j)}$ are depicted in Figs. \ref{plot_node_lowest_degree_growth_vs_connection_ink}, \ref{plot_normal_node_highest_degree_growth_vs_connection_ink} and \ref{plot_power_law_node_lowest_degree_growth_vs_connection_ink} respectively.
For all three distributions, the cummulative degree distribution follows the monomolecular distribution. The only difference for gaussian is that the starting value of the node with $min{(\Delta d_j)}$ also starts at low values with $0$ till 44 connections. All the six previous plots are a function of the number of connections, i.e., $k$. Hence, if the number of connections $k$ is broadcasted in the entire network then any node can estimate their quantum resource requirement following the curve fits.
\begin{figure}[ht]
\centering
\includegraphics[width=\columnwidth]{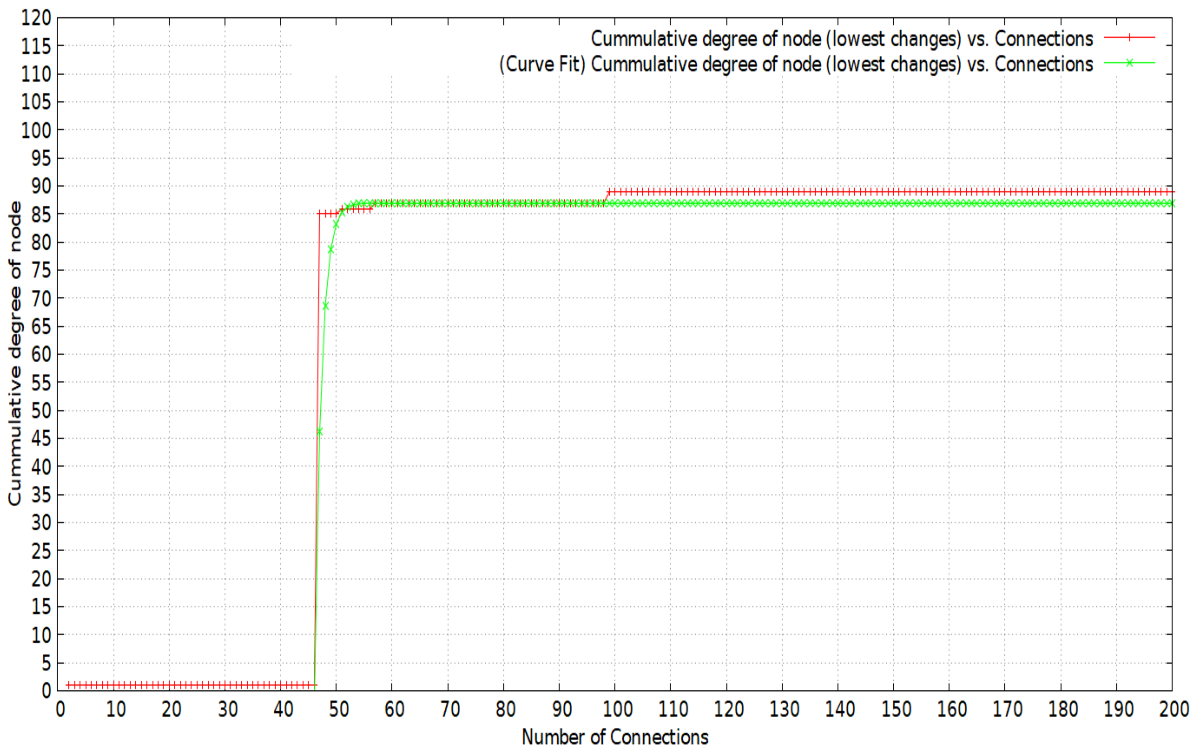}
\caption{Gaussian - Cummulative degree of the node with lowest changes vs. connection. Curve fit $d_j^{(k)} = 87-450e^{-0.80k}$ for $k>44$, $0$ otherwise}
\label{plot_normal_node_lowest_degree_growth_vs_connection_ink}
\end{figure}
\subsection{Discussion}
From the results above, several important inferences can be made which can help in providing an estimate of quantum resource allocation in the network. For all three connection setup distributions, namely, uniform, gaussian and power law, the usage frequencies of edges, both physical and virtual, in entangled graphs can be approximated with a power law function. Comparing all three distributions for the connection graph, the power law curve fit for edges in entangled graphs is more pronounced for power law and uniform scenarios. With an increased number of connections, the edge centralities of the entangled graph also get amplified. For gaussian distribution, the edge centralities (physical and virtual) get amplified with decreasing standard deviations. For power law distribution, similar behaviours are observed for decreasing exponents. Total cummulative entanglements with increasing connections follow patterns similar to monomolecular function. The degree distribution of each node shows a monomolecular growth with an increasing number of connections for all three distributions. The sum of changes in degrees with increasing connections shows two distinct levels partly due to the proactive entanglements for all three connection setup distributions. Both scenarios with the highest and lowest degree changes show monomolecular growths with different parameters.

\section{Conclusion and Future Work}\label{section_conclusion}
The paper studied the effects of centralities of entanglements due to different connection patterns. It is considered a quantum network model which consists of two graphs, namely, the entangled graph and the connection graph. The entangled graph contains the physical quantum links and those that are created by entanglement swaps. The connection graph is driven by three different usage patterns, namely, uniform, gaussian and power law. Both edge centralities (measured as the frequencies of edges used in the formation of the entangled graph) and node centralities (degree of nodes in the entangled graph) were investigated using simulations. Results showed that the edge centralities of the entangled graph showed power law distribution for most cases and the node centralities followed monomolecular distribution. Growth in entanglements with the number of connections also follows the monomolecular distribution. These results can help in the allocation of quantum resources. For example, quantum technologies with high reliability but low decoherence time may be deployed along the edges with high centralities since they are more likely to be used frequently. Also, virtual links with high centralities could be upgraded to physical edges as a deployment strategy. Node centralities give an estimate of the number of entanglements that will be handled as connections increase.

Future work will consider the tradeoff between proactive and reactive entanglements on centralities. Also, further studies are required to understand the impact on centralities following other approaches of entanglement distribution in the literature.

% or


%\appendices

%\section{}

\ifCLASSOPTIONcompsoc
  \section*{Acknowledgments}
\else
  \section*{Acknowledgment}
\fi

The authors would like to thank Mphasis NextLab for sponsoring this research.

\ifCLASSOPTIONcaptionsoff
  \newpage
\fi



%\bibliographystyle{IEEEtran}
%\begin{thebibliography}{1}

%H.~Kopka and P.~W. Daly, \emph{A Guide to \LaTeX}, 3rd~ed.\hskip 1em plus 0.5em minus 0.4em\relax Harlow, England: Addison-Wesley, 1999.

%\end{thebibliography}

\bibliographystyle{IEEEtran}
\bibliography{IEEEabrv,sample}

%\end{IEEEbiography}


%\newpage



%\vfill


\end{document}